\documentclass{iopjournal}
\renewcommand{\headrulewidth}{0pt}
\renewcommand{\footrulewidth}{0pt}

\usepackage{amssymb}
\usepackage{amsmath}
\usepackage{physics}
\usepackage{latexsym}

\usepackage{verbatim}
\usepackage{mathrsfs}
\usepackage{amsfonts}

\usepackage{graphicx}
\usepackage{epsfig}

\usepackage{units}
\usepackage{overpic}
\usepackage[utf8]{inputenc}
\usepackage{rotating}
\usepackage{enumerate}

\usepackage{soul}

\usepackage{xfrac}
\usepackage{xcolor}
\usepackage{multirow}

\usepackage{ragged2e}
\usepackage{afterpage}
\usepackage{cite}

\usepackage{placeins}

\justifying

\begin{document}


\title{Noble gravitational atoms: Self-gravitating black hole scalar wigs with angular momentum number}

\newcommand{\re}{\mbox{Re}}
\newcommand{\im}{\mbox{Im}}
\newcommand{\diag}{\mbox{diag}}
\newcommand{\Real}{\mathbb{R}}
\newcommand{\Complex}{\mathbb{C}}

\newcommand{\argelia}[1]{\textcolor{red}{{\bf Argelia: #1}}} 
\newcommand{\dario}[1]{\textcolor{red}{{\bf Dario: #1}}}  
\newcommand{\juanc}[1]{\textcolor{olive}{{\bf JC: #1}}}
\newcommand{\juan}[1]{\textcolor{cyan}{{\bf Juan B: #1}}}
\newcommand{\alberto}[1]{\textcolor{blue}{{\bf Alberto: #1}}}
\newcommand{\miguela}[1]{\textcolor{red}{{\bf MiguelA: #1}}}    
\newcommand{\mm}[1]{\textcolor{orange}{{\bf MM: #1}}}     
\newcommand{\OS}[1]{\textcolor{blue}{{\bf OS: #1}}}

\author{Miguel Alcubierre$^1$, Juan Barranco$^2$, Argelia Bernal$^2$, Juan Carlos Degollado$^3$, Alberto Diez-Tejedor$^2$, Miguel Megevand$^{4,*}$, Dar\a'io N\a'u\a~nez$^1$, Olivier Sarbach$^5$}

\affil{$^1$Instituto de Ciencias Nucleares, Universidad Nacional Aut\'onoma de M\'exico, Circuito Exterior C.U., A.P. 70-543, Coyoac\'an, M\'exico 04510, CdMx, M\'exico}

\affil{$^2$Departamento de Física, División de Ciencias e Ingenierías, Campus León, Universidad de Guanajuato, C.P. 37150, León, México}

\affil{$^3$Instituto de Ciencias F\'isicas, Universidad Nacional Aut\'onoma de M\'exico, Apdo. Postal 48-3, 62251, Cuernavaca, Morelos, M\'exico}

\affil{$^4$Instituto de F\'isica Enrique Gaviola, CONICET. Ciudad Universitaria, 5000 C\'ordoba, Argentina}

\affil{$^5$Instituto de F\'isica y Matem\'aticas, Universidad Michoacana de San Nicol\'as de Hidalgo, Edificio C-3, Ciudad Universitaria, 58040 Morelia, Michoac\'an, M\'exico}

\affil{$^*$Corresponding author. E-mail: mfmegevand@gmail.com}

\begin{center}
\today
\end{center}

\begin{abstract} \justifying \noindent
We present new spherically symmetric solutions of the Einstein-Klein-Gordon equations in a quasi-stationary approximation that describe self-gravitating scalar field configurations around a black hole, including angular momentum number $\ell$.
An approach analogous to the one which gives rise to $\ell$-boson stars is used here to construct self-gravitating ``gravitational atoms" with $\ell\ge0$.
We refer to these new solutions as {\it noble gravitational atoms}, by analogy with noble atoms, which are characterized by closed electron shells. 
We show that, in the proper limit, noble gravitational atoms approach $\ell$-boson stars globally, displaying noticeable differences only in a region very close to the event horizon.
Noble gravitational atoms with $\ell>0$ sometimes present density maxima located at relatively large radii, with small  density close to the horizon for $\ell>1$. Furthermore, they do not always present the typical density spike at the event horizon if $\ell > 0$; on the contrary, they sometimes exhibit a small dip there. When $\ell=0$, a spike can appear, but its contribution to the total mass density is always negligible. 
The size, density, and lifetime of these objects vary significantly depending on the parameters, being in some cases as large as galaxies, as dilute as dark matter, and as long-lived as the Universe itself.
\end{abstract}

\section{Introduction}

The distribution of baryons and dark matter in the center of most galaxies can be strongly influenced by the presence of supermassive black holes~\cite{Bahcall:1976aa, Young1980, Quinlan:1994ed, Gondolo:1999ef, Sadeghian:2013laa, Ferrer:2017xwm}. As shown in previous work~\cite{Gondolo:1999ef, Sadeghian:2013laa, Ferrer:2017xwm}, if the dark matter has had sufficient time to relax dynamically, the adiabatic growth of the central black hole can steepen the inner dark matter profile and produce a characteristic density ``spike''. This spike increases the dark matter density near the black hole, which can in turn modify nearby stellar orbits and, in annihilating or decaying dark matter models, leads to potentially detectable fluxes of high energy cosmic rays from galactic centers~\cite{Bertone:2004pz, Bergstrom:1997fj, Chan:2019ptd, Barman:2022jdg, Johnson:2019hsm}.

In the late 90's and at the beginning of this century, the idea of a fundamental spin-zero particle that plays the role of dark matter with an ultralight mass of the order of $\mu \sim 10^{-22}$ eV 
was proposed~\cite{Hu:2000ke, Matos:2000ng, Matos:2000ss,Matos:2000ki,Marsh:2015xka,Hui:2016ltb,2019FrASS...6...47U,Ferreira:2020fam,Hui:2021tkt,matos2023short}. 
In some of these models, the dark matter particle has no self-interactions or couplings to the standard model, interacting only via minimal coupling to gravity, as we assume here. 
These extremely small dark matter masses are associated with parsec-size Compton wavelengths, much larger than the observed black holes.
Thus, in ultralight dark matter models, the physical processes operating near black holes are very different from those in the standard cold dark matter scenario, and present potential implications for observable signatures.

We have previously studied quasi-stationary scalar field distributions around a Schwarzschild black hole in the test field approximation~\cite{Barranco:2011eyw, Barranco:2012qs, Barranco:2013rua}, finding configurations that can last for cosmological times.
Later, we extended our study to self-gravitating scalar fields~\cite{Barranco:2017aes, Alcubierre:2024mtq}, with similar results regarding the existence of very long-lived solutions.
Furthermore, these solutions were found to have almost the same density profiles as boson stars at scales much larger than the black hole's horizon.
As shown in~\cite{Alcubierre:2024mtq}, these more realistic models are even adequate for describing
the galactic dark matter  core of the Milky Way, which hosts Sgr.~A* at its center~\cite{Alcubierre:2024mtq}.
Such scenarios have also been studied dynamically in~\cite{Annulli:2020lyc, Cardoso:2022nzc}  with similar conclusions, where the authors also identified the existence of a quasi-adiabatic regime that they have confirmed through time evolutions.
The spectrum of these objects has been studied, both in the test-field approximation and in the self-gravitating case, for instance, in~\cite{Barranco:2013rua} and~\cite{DellaRocca:2025xwz}.
For a rotating black hole, 
the scalar field can form a large cloud by extracting energy from the black hole, 
in a process called super-radiance~\cite{Zeldovich, Zeldovich72, Press:1972zz, penrose, Starobinsky:1973aij, Starobinskii:1973hgd, Detweiler:1980uk, Brito:2014wla}. 
The black hole's rotation and the frequency of the scalar field can synchronize in this case, and stationary distributions emerge that constitute Kerr black hole hair~\cite{Herdeiro:2014goa, Herdeiro:2015gia, Cunha:2015yba}.

In this work, we construct self-gravitating quasi-stationary scalar configurations around a black hole in spherical symmetry, including angular momentum number $\ell$. 
The construction is based on a method implemented in e.g.~\cite{Olabarrieta:2007di} and explored further in~\cite{Alcubierre:2018ahf, Alcubierre:2019qnh, Alcubierre:2021mvs, Alcubierre:2021psa, Jaramillo:2020rsv, Jaramillo:2022zwg, Alcubierre:2022rgp} for $\ell$-boson stars. 
In these cases, $2\ell+1$ classical fields with equal amplitude and angular momentum number $\ell$ produce a spherically symmetric stress-energy tensor. 
Hence, a distinctive feature of these configurations is that the spacetime geometry is spherically symmetric, while the scalar fields themselves are not.

When considering a quasi-stationary system that includes a black hole in its center and is characterized by an angular momentum number $\ell$, we call the resulting solutions {\it noble gravitational atoms}, because they are spherically symmetric and exhibit closed shells. 
Furthermore, in the test-field approximation, their energy analogs, $\omega_n^2$, characterized by the principal quantum number $n$, resemble the hydrogen energy levels when $\ell=0$, i.e, $\omega_n^2 \sim 1/n^2$.
The analogy, of course, is not perfect. 
For instance, when $\ell\neq 0$, the frequencies $\omega_{n\ell}$ cease to match the hydrogenic spectrum, and there are no stationary solutions for massive scalar fields in the vicinity of Schwarzschild black holes. 
Nevertheless, for small values of the ``gravitational fine structure constant'', defined in terms of the black hole and scalar field masses, $M$ and $\mu$, and the gravitational constant $G$, as $\alpha_G=GM\mu$, 
scalar fields can survive longer than the age of the Universe, as it was already shown, for the case of a single scalar field, in~\cite{Barranco:2011eyw, Barranco:2012qs, Barranco:2013rua, Barranco:2017aes, Alcubierre:2024mtq,Annulli:2020lyc, Cardoso:2022nzc}.
For this reason, we have previously called those solutions ``wigs'', as they resemble, but are not actually, black hole hair. 

The present study naturally extends the construction of a single self-gravitating scalar field surrounding black holes by considering $2\ell+1$ scalar fields with the same radial profile. 
We show that, within a certain range of the parameter space, the corresponding generalization leads to configurations that are practically indistinguishable from $\ell$-boson stars except for differences localized near the black hole. 

Finally, we would like to address the relevance of considering such a particular combination of $2\ell+1$ classical fields, as this choice might seem arbitrary {\it a priori}. 
As we have previously shown, at least in the case of $\ell$-boson stars, a stationary, spherically symmetric configuration with $2\ell+1$ effective classical fields emerges naturally within the context of the semiclassical theory of gravity with a {\em single} quantum scalar field~\cite{Alcubierre:2022rgp}.
Furthermore, as shown in~\cite{Roque:2023sjl, Nambo:2023yut}, in the non-relativistic limit $\ell$-boson stars with $\ell=0$ and $\ell=1$ are linearly stable.
Although the case of interest here, in which a black hole is present, may be different, these results 
encourage the consideration, as physically plausible, of other types of spherically symmetric objects consisting of $2\ell+1$ non-symmetric scalar fields, like the noble gravitational atoms. 

The article is organized as follows. We begin, in section~\ref{s:formulations}, by presenting our formulation, which generalizes our previous work~\cite{Alcubierre:2024mtq} to arbitrary values of $\ell$, and obtain the main equations and boundary conditions that give rise to the noble gravitational atoms. Besides including the contribution from the angular momentum number~$\ell$, we also provide a detailed discussion of the quasi-stationary approximation, expanding the one in our previous work to arbitrary $\ell\geq 0$.
In section~\ref{s:results}, we present and analyze our solutions. Note that some results for the particular case when $\ell=0$ were already studied in~\cite{Alcubierre:2024mtq}. 
Nevertheless, we will include them in our analysis, first, to further expand on the results presented in that short letter, and second, to serve as a reference to compare with the $\ell>0$ cases.
Finally, in section~\ref{s:conclusions}, we offer some final remarks. Further examples of noble gravitational atoms and details regarding dark matter spikes are included in appendixes. Throughout this work, we use Planck units, such that $G=c=\hbar=1$, and choose the metric signature $(-, +, +, +)$.

\section{Self-gravitating quasi-stationary model}
\label{s:formulations}

In this section, we derive the system that describes the self-gravitating configurations in the quasi-stationary approximation that are relevant to this article. As we will see, this system consists of a coupled set of nonlinear ordinary differential equations, together with appropriate boundary conditions at the horizon and at infinity, which leads to a nonlinear eigenvalue problem for the complex frequency $s$ of the scalar field. This system is given by equations~(\ref{Eq:eqns}) below with boundary conditions~\eqref{boundary_conditions} and~\eqref{rightboundary}.
As we show, the solution of this system provides \emph{exact} initial data for the Einstein-Klein-Gordon system with multiple fields and \emph{approximate} time-dependent solutions of these equations which are expected to be valid on time scales that are small when compared to the decay time $t_0 = 1/|\re(s)|$ and the accretion time $t_{a}$, defined in Eq.~\eqref{taccr} below.

\subsection{The generalized Eddington-Finkelstein gauge} 
\label{s:formulation}

Our method relies crucially on the use of horizon-penetrating coordinates, which are regular at the horizon. Following~\cite{Alcubierre:2024mtq}, we write the metric in the form
\begin{equation}
ds^2 = -a(t,r)^2 dt^2 + dr^2 + \frac{2m(t,r)}{r}\left( a(t,r) dt + dr \right)^2 + r^2 d\Omega^2,
\label{Eq:MetricEFGauge}
\end{equation}
where $a(t,r)$ and $m(t,r)$ are positive functions depending on the time $t$ and radial areal coordinate $r$, and $d\Omega^2$ denotes the metric on the unit two-sphere. Here, the function $m(t,r)$ describes the Misner-Sharp or Hawking mass (with respect to the spheres of constant $t$ and $r$), and as described in~\cite{Alcubierre:2024mtq} the surface $r = 2m$ describes a non-expanding horizon. For the following, it is convenient to introduce the vector fields
\begin{equation}
n^\mu\partial_\mu := \frac{\gamma}{a}\frac{\partial}{\partial t} - \frac{2m}{r}\frac{1}{\gamma}\frac{\partial}{\partial r},\qquad
w^\mu\partial_\mu := \frac{1}{\gamma}\frac{\partial}{\partial r},
\end{equation}
where $\gamma:=\sqrt{1 + 2m/r}$. The first one is the future-directed unit normal vector field to the $t={\rm const.}$ hypersurfaces. The second one is unitary, orthogonal to $n$, and points in the radial outward direction.

Einstein's field equations for the metric~(\ref{Eq:MetricEFGauge}) reduce to
\begin{subequations}
\label{Eq:D0primeEF}
\begin{align}
D_0 m &= \frac{\kappa r^2}{2\gamma}
\left[ T(n,w) + \frac{2m}{r} 
T(w,w) \right],
\label{Eq:D0mEF}\\
m' &= \frac{\kappa r^2}{2}
\left[ T(n,n) + \frac{2m}{r} T(n,w) \right],
\label{Eq:mprimeEF}\\
\frac{a'}{a} &= \frac{\kappa r}{2\gamma^2}T(n-w,n-w),
\label{Eq:aprimeEF}
\end{align}
\end{subequations}
with $D_0 := n^\mu\partial_\mu$, the prime denoting a partial derivative with respect to $r$, $\kappa = 8\pi$  is the gravitational coupling constant, and where we have abbreviated $T(X,Y) := T_{\mu\nu} X^\mu Y^\nu$. Note that when the stress-energy tensor $T_{\mu\nu}$ vanishes, one obtains directly $m = {\rm const.}$ and that $a$ is a function of time only, thus recovering the Schwarzschild solution in Kerr-Schild form.

The three-metric and extrinsic curvature associated with the $t= {\rm const.}$ hypersurfaces are given by
\begin{subequations}
\label{Eq:InitialData}
\begin{align}
h_{ij} dx^i dx^j &= \gamma^2 dr^2 + r^2 d\Omega^2,
\label{Eq:ThreeMetric}\\
k_{ij} dx^i dx^j &= \left[ \left( 1 + \frac{m}{r} \right)\left( \frac{2m}{r^2} - \kappa r T(n,n) \right) + \frac{\kappa r}{2} T(n,w) \right]\frac{dr^2}{\gamma} - \frac{2m}{\gamma} d\Omega^2,
\label{Eq:ExtrinsicCurvature}
\end{align}
\end{subequations}
where we have used equations~\eqref{Eq:D0primeEF} to eliminate $D_0 m$, $m'$, and $a'$.
One can verify that the fulfillment of equation~(\ref{Eq:mprimeEF}) implies that these expressions automatically satisfy both the Hamiltonian and the momentum constraints, which in spherical symmetry reduce to
\begin{subequations}
\begin{align}
& \frac{R}{2} + (2k^r{}_r + k^\vartheta{}_\vartheta)k^\vartheta{}_\vartheta = \kappa T(n,n),\\
& k^r{}_r - (r k^\vartheta{}_\vartheta)' = \frac{\kappa r}{2}\gamma T(n,w),
\end{align}
\end{subequations}
with $R = 4r^{-2}\gamma^{-4}( m' + 2m^2/r^2)$ the Ricci scalar associated with the three-metric~(\ref{Eq:ThreeMetric}).

\subsection{The \texorpdfstring{$\ell$}{l}-boson star ansatz} 
\label{s:ellansatz}

For the matter content, we consider a family of $2\ell+1$ classical complex scalar fields $\Phi_m$, each with the same mass $\mu$, which gives rise to the stress-energy tensor\footnote{{  Although some forms of interactions could in principle be included in our model, we only consider free scalar fields in this work. Self-interactions may be relevant for axion-like candidates; however, they are generically controlled by the axion mass and are therefore extremely weak in the ultralight case. Moreover, such interactions involve higher powers of the axion field in the equations of motion. Since the scalar field remains bounded in our configurations, there always exists a region of the solution space for which the self-interaction effects remain negligible. 
For these reasons we neglect self-interaction terms in our work.}}
\begin{equation}
T_{\mu\nu} = \sum\limits_{m=-\ell}^\ell\left[ 
\nabla_{(\mu}\Phi_m^*\cdot\nabla_{\nu)}\Phi_m 
 - \frac{1}{2} g_{\mu\nu}\left( \nabla^\alpha\Phi_m^*\cdot\nabla_\alpha\Phi_m + \mu^2|\Phi_m|^2 \right) \right],
\end{equation}
where $\nabla_\mu$ denotes the covariant derivative with respect to the space-time metric.
This is the same matter content than for the $\ell$-boson stars~\cite{Alcubierre:2018ahf, Alcubierre:2019qnh, Alcubierre:2021mvs, Alcubierre:2021psa, Jaramillo:2020rsv, Jaramillo:2022zwg, Alcubierre:2022rgp}, although in the present article we are interested in the case where these scalar fields coexist with a non-rotating black hole. To proceed, we make the $\ell$-boson star ansatz 
\begin{equation}
\Phi_m = \phi_\ell(t,r) Y^{\ell m}(\vartheta,\varphi),
\end{equation}
with $Y^{\ell m}$ the standard spherical harmonics, and $\phi_\ell$ a function which is independent of $m$. 

Using the expressions~(A.8) from~\cite{Alcubierre:2018ahf}, one obtains
\begin{subequations}
\label{Eq:Tmunu}
\begin{align}
T(n,n) &= \frac{2\ell+1}{8\pi}
\left( |\pi_\ell|^2 + |\chi_\ell|^2 + \mu_\ell^2|\phi_\ell|^2 \right),
\\
T(w,w) &= \frac{2\ell+1}{8\pi}
\left( |\pi_\ell|^2 + |\chi_\ell|^2 - \mu_\ell^2|\phi_\ell|^2 \right),
\\
T(n,w) &= \frac{2\ell+1}{4\pi}\re( \pi_\ell^*\chi_\ell ), 
\end{align}
where we have defined $\pi_\ell := D_0\phi_\ell$ and $\chi_\ell := \gamma^{-1}\phi'_\ell$ and introduced the shorthand notation 
$\mu_\ell^2 := \mu^2 + \ell(\ell+1)/r^2$. 
Note that
\begin{equation}
T(n-w,n-w) = \frac{2\ell+1}{4\pi}\left| \pi_\ell - \chi_\ell \right|^2.
\end{equation}
\end{subequations}

The Klein-Gordon equations, $-\nabla^\alpha\nabla_\alpha\Phi_m + \mu^2\Phi_m = 0$, yield the evolution system 
for $(\phi_\ell,\pi_\ell)$. 
In the generalized Eddington-Finkelstein gauge, the field equations reduce to:
\begin{subequations}
\label{Eq:D0}
\begin{align}
D_0\phi_\ell &= \pi_\ell,
\label{Eq:PhiEF}\\
D_0\pi_\ell &= \frac{1}{a r^2}\left( r^2\frac{a}{\gamma^2}\phi_\ell' \right)' 
 - 3c\pi_\ell - \mu_\ell^2\phi_\ell,
\label{Eq:PiEF}
\end{align}
where $3c$ denotes the trace of the extrinsic curvature,
\begin{equation}
3c = -\frac{1}{\gamma^3}\left( 1 + \frac{3m}{r} \right)\frac{2m}{r^2}
 + \frac{\kappa r}{2\gamma^3}
 \left[ T(n,w) - 2\left( 1 + \frac{m}{r} \right) T(n,n) \right],
\label{Eq:3cEF}
\end{equation}
\end{subequations}
and the functions $m$ and $a$ are determined by equations~\eqref{Eq:D0primeEF}.

Initial data for $(\phi_\ell,\pi_\ell)$ on a $t={\rm const.}$ hypersurface can be freely specified. They determine the relevant components~\eqref{Eq:Tmunu} of the stress-energy tensor, up to the mass function $m$ which appears in the factor $\gamma$ in $\chi_\ell = \gamma^{-1}\phi_\ell'$. In turn, $m$ can be obtained by integrating equation~\eqref{Eq:mprimeEF}. This allows one to define the three-metric and extrinsic curvature by means of equations~\eqref{Eq:InitialData}, and by construction they automatically satisfy the constraints. In what follows, we describe a method for obtaining exact initial data giving rise to {approximate} quasi-stationary solutions of the evolution system~\eqref{Eq:D0primeEF} and \eqref{Eq:D0}.

\subsection{The quasi-stationary approximation} 
\label{s:QS}

Next, we formulate our proposal for constructing approximate quasi-bound states in the self-gravitating case. It is based on the idea that in the test field limit there exist solutions where $\phi_\ell$ and $\pi_\ell$ both have the time-dependency of $e^{st}$ with $s=\sigma+i\omega$ complex, the real part $\sigma$ being negative and describing the time-decay. In the self-gravitating case, one cannot expect to encounter solutions of the exact same form, since the metric is time-dependent. However, assuming that the metric varies in time much slower than the absolute values of $\phi_\ell$ and $\psi_\ell$, it is plausible to expect an approximate solution of the mode form,
\begin{equation}
\left( \begin{array}{c} \phi_\ell(t,r) \\ \pi_\ell(t,r) \end{array} \right)
 = e^{st} \left( \begin{array}{c} \psi_\ell(r) \\ \theta_\ell(r) \end{array} \right),
\label{Eq:ModeAnsatz}
\end{equation}
with purely radial functions $\psi_\ell(r)$ and $\theta_\ell(r)$.
Introducing this ansatz into equations~\eqref{Eq:D0}
one obtains
\begin{subequations}
\begin{align}
\left( \gamma\frac{s}{a} - \frac{2m}{r}\frac{1}{\gamma} \frac{d}{dr} \right)\psi_\ell 
 &= \theta_\ell,
 \label{Eq:theta}\\
\left( \gamma\frac{s}{a} - \frac{2m}{r}\frac{1}{\gamma} \frac{d}{dr} \right)\theta_\ell 
 &= \frac{1}{a r^2}\frac{d}{dr} \left( r^2\frac{a}{\gamma^2} \psi_\ell' \right) 
 - 3c\theta_\ell - \mu_\ell^2\psi_\ell,
\end{align}
\end{subequations}
where
$3c$ is given by the expression~(\ref{Eq:3cEF}) and $m$ and $a$ can be determined by integrating equations~(\ref{Eq:mprimeEF},~\ref{Eq:aprimeEF}). Further, in these equations, the components of the stress-energy tensor are
\begin{subequations}
\label{Eq:Txx}
\begin{align}
T(n,n) &= \frac{2\ell+1}{8\pi} e^{2\sigma t}
 \left( |\theta_\ell|^2 + \gamma^{-2}|\psi_\ell'|^2 + \mu_\ell^2|\psi_\ell|^2 \right),
 \label{Eq:Tnn}\\
T(w,w) &= \frac{2\ell+1}{8\pi} e^{2\sigma t}
\left( |\theta_\ell|^2 + \gamma^{-2}|\psi_\ell'|^2 - \mu_\ell^2|\psi_\ell|^2 \right),
 \label{Eq:Tww}\\
T(n,w) &= \frac{2\ell+1}{4\pi} e^{2\sigma t}\frac{1}{\gamma}\re(\theta_\ell^*\psi_\ell'),
 \label{Eq:Tnw}\\
T(n-w,n-w) &= \frac{2\ell+1}{4\pi} e^{2\sigma t}\left| \theta_\ell - \frac{1}{\gamma}\psi_\ell' \right|^2.
\end{align}
\end{subequations}

Because of the factors $e^{2\sigma t}$, one cannot assume that the fields are strictly time-independent, as anticipated above. However, one can always look for solutions $(\psi_\ell,\theta_\ell,m,a)$ of the above system at the initial time $t = 0$, where $e^{2\sigma t} = 1$. Since (for the situations we are interested in) we expect $|\sigma|$ to be very small, the resulting configurations should yield a good approximation to the true time-dependent solution for time scales $t\ll 1/|\sigma|$.\footnote{As we will see shortly, another limitation comes from the accretion time scale.} Furthermore, the configurations obtained in this way can be used as initial data for the full system, which can then be evolved using the full (time-dependent) equations without approximations.

Therefore, at $t=0$, our radial system of equations describing the self-gravitating quasi-bound states in the quasi-stationary approximation can be written as follows. Eliminating $\theta_\ell$, the scalar field equation takes the form
\begin{equation}
-\left(1 - \frac{2m}{r} \right)\psi_\ell'' + B\psi_\ell' + C\psi_\ell = 0,
\label{Eq:Psi}
\end{equation}
where the coefficients $B$ and $C$ are defined by
\begin{align}
\begin{split}
B &= -\frac{1}{a r^2}\frac{d}{dr} \left( r^2\frac{a}{\gamma^2} \right)
 + \frac{2m}{r\gamma}\left[ \frac{d}{dr}\left( \frac{2m}{r\gamma} \right) - 3c - 2\gamma\frac{s}{a} \right],\\
C &= \mu_\ell^2 + \gamma^2\left( \frac{s}{a} \right)^2
 + \left[ 3\gamma c - \frac{2m}{r}\frac{a}{\gamma}
\frac{d}{dr}\left( \frac{\gamma}{a} \right) \right] \frac{s}{a},
\end{split}
\end{align}
and the metric coefficients are obtained by integrating
\begin{align}
m' &= \frac{\kappa r^2}{2}
\left[ T(n,n) + \frac{2m}{r} T(n,w) \right],
\label{Eq:mprimeEFBis}\\
\frac{a'}{a} &= \frac{\kappa r}{2\gamma^2} T(n-w,n-w).
\label{Eq:aprimeEFBis}
\end{align}
Using the last two equations and the expression~(\ref{Eq:3cEF}) for $3c$ one can represent the coefficients $B$ and $C$ in a more explicit way:
\begin{align}
\begin{split}
B &= -\frac{2}{r}\left( 1 - \frac{m}{r} \right) - \frac{4m}{r}\frac{s}{a}
 + \frac{\kappa r}{2\gamma^4}\left[ (2\gamma^4-1) T(n,n)
  + 2\left( 1 + \frac{m}{r}\gamma^4 \right)T(n,w) - T(w,w) \right],\\
C &= \mu_\ell^2 + \gamma^2\left( \frac{s}{a} \right)^2 - \frac{2m}{r^2}\frac{s}{a}
 + \frac{\kappa r}{2\gamma^2}\left[ -2\left( 1 + \frac{m}{r} \right)T(n,n) 
  + (2 - \gamma^4)T(n,w) + \frac{2m}{r} T(w,w) \right] \frac{s}{a}. 
\end{split}
\end{align}
Using the expressions~\eqref{Eq:Txx},
in which $\theta_\ell$ can be eliminated using equation~(\ref{Eq:theta}), we arrive at the final system
\begin{subequations}
\label{Eq:eqns}
\begin{align}
\left(1 - \frac{2m}{r} \right)\psi_\ell'' &= B\psi_\ell' + C\psi_\ell,
\label{Eq:PsiBis}\\
2m' &= \kappa_\ell r^2 e^{2\sigma t}\left[ \gamma^2 \left| \frac{s}{a}\psi_\ell \right|^2 
 + \mu_\ell^2 |\psi_\ell|^2 + \left(1 - \frac{2m}{r} \right) \left| \psi_\ell' \right|^2 \right],
\label{Eq:mBis}\\
\frac{a'}{a} &= \kappa_\ell r e^{2\sigma t}\left| \frac{s}{a}\psi_\ell - \psi_\ell' \right|^2,
\label{Eq:aBis}
\end{align}
\end{subequations}
where we have abbreviated $\kappa_\ell := (2\ell+1)\kappa/(8\pi)$ and where
\begin{align}
\begin{split}
B &= -\frac{2}{r}\left( 1 - \frac{m}{r} \right) - \frac{4m}{r}\frac{s}{a}
\\
 &+ \frac{\kappa_\ell r}{\gamma^2}\left[ 
  (\gamma^4-1)\left| \frac{s}{a}\psi_\ell \right|^2 + \gamma^2\mu_\ell^2 |\psi_\ell|^2
   + (1+\gamma^2)\left( 1 - \frac{2m}{r} \right)\re\left( \frac{s^*}{a}\psi_\ell^*\psi_\ell' \right)
  \right],
\\
C &= \mu_\ell^2 + \gamma^2\left( \frac{s}{a} \right)^2 - \frac{2m}{r^2}\frac{s}{a}
\\
 &-\kappa_\ell r\left[ \left| \frac{s}{a}\psi_\ell \right|^2 + \mu_\ell^2 |\psi_\ell|^2
   + \left( 1 - \frac{2m}{r} \right)\left| \psi_\ell' \right|^2
   - \left( 1 - \frac{2m}{r} \right)\re\left( \frac{s^*}{a}\psi_\ell^*\psi_\ell' \right)
\right] \frac{s}{a}. 
\end{split}
\end{align}
Note that the expressions for $B$ and $C$ and the right-hand side of equation~(\ref{Eq:mBis}) are independent of $\psi_\ell'$ when $r = 2m$.

The system~\eqref{Eq:eqns} for $(\psi_\ell,m,a)$ constitutes a nonlinear eigenvalue problem for the complex frequency $s$. Note the invariance with respect to the rescaling $(a,s)\mapsto (\lambda a,\lambda s)$. The correct boundary conditions at the horizon and the asymptotic region $r\to \infty$ are determined in the next subsection.
Finally, we define, for later use, the energy density, $\rho:=m^\prime /(4\pi r^2)$, which can be evaluated from the right-hand side of equation~\eqref{Eq:mBis}.

\subsection{Boundary conditions}
\label{s:BC}

We first discuss inner boundary conditions at an apparent horizon, where $2m = r$. A difficulty arises because the system~\eqref{Eq:eqns}
is singular at such points, since the coefficient in front of $\psi_\ell''$ in equation~(\ref{Eq:PsiBis}) vanishes there.

To get an idea of what occurs in the vicinity of $2m = r$, we first turn off the self-gravity of the field by setting $\kappa=0$. This yields $m = {\rm const.}$ and $a = {\rm const.} := 1$, and the coefficients in equation~(\ref{Eq:PsiBis}) reduce to
\begin{align}
\begin{split}
B &= -\frac{2}{r}\left( 1 - \frac{m}{r} \right) - \frac{4m}{r} s,\\
C &= \mu_\ell^2 + \gamma^2 s^2 - \frac{2m}{r^2} s.
\end{split}
\end{align}
Introducing the new coordinate $z := r/(2m) - 1$, which is zero at the horizon, equation~(\ref{Eq:PsiBis}) is transformed into
\begin{equation}
z\frac{d^2\psi_\ell}{dz^2} + \Delta(z)\frac{d\psi_\ell}{dz} = \Gamma(z)\psi_\ell,
\label{Eq:PsiInTermsOfz}
\end{equation}
with the coefficients
\begin{align}
\begin{split}
\Delta(z) &:= -rB = 1 + 4ms + \frac{z}{1+z},\\
\Gamma(z) &:= 2m r C =  (2m\mu)^2(1+z) + \frac{\ell(\ell+1)}{1+z}
 + (2ms)^2(2+z) - \frac{2ms}{1+z}.
\end{split}
\end{align}

Since $\Delta$ and $\Gamma$ are analytic functions of $z$ in an open neighborhood of the region of interest $z\geq 0$, it follows that equation~\eqref{Eq:PsiInTermsOfz} has a regular singular point at $z=0$ with local solutions behaving as $z^\lambda$ with $\lambda$ determined by the characteristic equation
\begin{equation}
\lambda(\lambda-1) + \Delta_0\lambda = 0,
\end{equation}
where $\Delta_0 := \Delta(0) = 1 + 4ms$. Therefore, there is a solution which is regular at $z=0$ and another one which behaves as $z^{-4ms}$ whose derivative diverges at $z=0$ and should therefore be discarded. 
The regular solution's first and second derivatives at $z=0$ can be obtained by evaluating equation~\eqref{Eq:PsiInTermsOfz} and its first derivative at $z=0$, which yields
\begin{equation}
\left. \frac{1}{\psi_\ell}\frac{d\psi_\ell}{dz} \right|_{z=0}
 = \frac{\Gamma_0}{\Delta_0},\qquad
\left. \frac{1}{\psi_\ell}\frac{d^2\psi_\ell}{d^2z} \right|_{z=0}
 = \frac{\Gamma_0^2 - \Gamma_0\Delta_1 + \Gamma_1\Delta_0}{\Delta_0(1+\Delta_0)},
\label{Eq:PsiFirstTwoDerivs}
\end{equation}
with\footnote{As we will see, the solution of Eq.~\eqref{Eq:PsiInTermsOfz} which is regular at the horizon can be represented in terms of the confluent Heun function, and Eq.~\eqref{Eq:PsiFirstTwoDerivs} gives the first two terms from the Taylor expansion at $z=0$ of these functions.}
\begin{align}
\begin{split}
\Delta_0 &= 1 + 4ms,\\
\Delta_1 &= 1,\\
\Gamma_0 &= (2m\mu)^2 + \ell(\ell+1) + 2(2ms)^2 - 2ms,\\
\Gamma_1 &= (2m\mu)^2 - \ell(\ell+1) + (2ms)^2 + 2ms.
\end{split}
\end{align}

Next, let us analyze the self-gravitating case. Here we introduce the parameter $r_0 := 2M > 0$, which is the radius of the apparent horizon, and write
\begin{align}
\begin{split}
r &= r_0(1 + z),\\
2m &= r_0\left[ 1 + z p(z) \right],
\end{split}
\end{align}
with $p(z)$ a positive function. Note that
\begin{equation}
1 - \frac{2m}{r} = \frac{z}{1+z}(1-p);
\end{equation}
hence, $p < 1$ is required to avoid a second apparent horizon. Using this, Eq.~(\ref{Eq:PsiBis}) can be brought again into the form~(\ref{Eq:PsiInTermsOfz}) where now the functions $\Delta$ and $\Gamma$ are defined by
\begin{equation}
\Delta(z) := -\frac{rB}{1-p},\qquad
\Gamma(z) := \frac{r_0 r C}{1-p}.
\end{equation}
If a regular solution at $z=0$ exists, then their first two derivatives at $z=0$ are given by equations~(\ref{Eq:PsiFirstTwoDerivs}), with $\Gamma_k$ and $\Delta_k$ denoting the $k'$-th order derivative of $\Gamma$ and $\Delta$ with respect to $z$, evaluated at $z=0$. 
However, in contrast to the test field limit, the functions $\Delta$ and $\Gamma$ now depend themselves on $\psi_\ell$ and $\psi_\ell'$. Hence, one would expect, a priori, equations~\eqref{Eq:PsiFirstTwoDerivs} to lead to implicit relations for the first two derivatives of $\psi_\ell$. Fortunately, such a complication does not arise. 
This is because in the expressions for $B$, $C$, and $2m'$, the coefficients multiplying $\psi_\ell'$, are zero when $z = 0$. 
Assuming, without loss of generality, that $a(r_0) = 1$, one finds
\begin{subequations}
\label{Eq:DeltaGamma0}
\begin{align}
p_0 &= \kappa_\ell\left[ 2\left| r_0 s \right|^2 + (\mu r_0)^2 + \ell(\ell+1) \right] |A|^2,
\label{Eq:p0}\\
\Delta_0 &= 1 + \frac{1}{1-p_0}
\left[ 2r_0 s + \frac{\kappa_\ell}{2} | r_0 s |^2 |A|^2 \right],
\label{Eq:Delta0}\\
\begin{split}
\Gamma_0 &= \frac{1}{1-p_0}\left[ (\mu r_0)^2 + \ell(\ell+1) + 2(r_0 s)^2 - (r_0 s)  -
  \left( p_0 - \kappa_\ell\left| r_0 s \right|^2 |A|^2 \right) (r_0 s) \right],
\end{split}
\label{Eq:Gamma0}
\end{align}
\end{subequations}
where the coefficient $p_0 := p(0)$ is equal to $2m'$ evaluated at $r = r_0$ and has been computed from equation~\eqref{Eq:mBis}. Note also that $A:=\psi_\ell(r_0)$ should be chosen small enough such that $p_0 < 1$.

In conclusion, the boundary conditions at the apparent horizon are: 
\begin{align}
\begin{split}
\psi_\ell(r_0) &= A,\\
\psi_\ell'(r_0) &= \frac{A}{r_0}\frac{\Gamma_0}{\Delta_0},\\
\psi_\ell''(r_0) &= \frac{A}{r_0^2}\frac{\Gamma_0^2 - \Gamma_0\Delta_1 + \Gamma_1\Delta_0}{\Delta_0(1+\Delta_0)},\\
2m(r_0) &= r_0,\\
a(r_0) &= 1,
\end{split}
\label{boundary_conditions}
\end{align}
where $\Gamma_0$ and $\Delta_0$ are given in equations~\eqref{Eq:DeltaGamma0} and the coefficients $\Gamma_1$ and $\Delta_1$ can be obtained by differentiating equation~(\ref{Eq:PsiInTermsOfz}) with respect to $z$. Although this calculation can be carried out (see~\cite{Alcubierre:2024mtq} for the case $\ell=0$), the numerical method described in the next section will not require explicit knowledge of the second derivative of $\psi_\ell$ at $r=r_0$, and for this reason we omit it. Besides $r_0 = 2M > 0$, the only free parameter at the horizon is the scalar field's amplitude $A$. We have assumed that $a(r_0) = 1$ to simplify the calculations; however, one can recover the general case by simply replacing $s$ with $s/a(r_0)$ in the expressions above. Note that the boundary condition for $\psi_\ell'/\psi_\ell$ differs from the test field case.
From equations~(\ref{Eq:mBis}, \ref{Eq:aBis}) one also obtains 
\begin{align}
\begin{split}
2m'(r_0) &= p_0,\\
a'(r_0) &= \frac{\kappa_\ell}{r_0}\left| r_0 s - \frac{\Gamma_0}{\Delta_0} \right|^2 |A|^2,
\end{split}
\end{align}
where $p_0$ is given in equation~\eqref{Eq:p0}.

Regarding the boundary conditions at infinity, assuming $a(r)\sim m(r)\sim {\rm const.}$ at large $r$, and choosing the decaying solution for $\psi(r)$, one obtains at the right boundary,  $r=r_1\gg r_0$, 
\begin{equation}
\begin{split}
\begin{gathered}
    \psi_\ell^\prime(r_1) = \left( - k + \frac{c}{r_1} \right) \psi(r_1), \label{rightboundary}\\
    k = \sqrt{\mu^2+s^2} ,\quad c = \left( \frac{\mu^2}{k} -2s -2k \right) m(r_1) - 1 . 
\end{gathered}
\end{split}
\end{equation}

Numerical solutions for the system~\eqref{Eq:eqns} (with $e^{2\sigma t} = 1$)
for $(\psi_\ell,m,a)$ with boundary conditions (\ref{boundary_conditions}, \ref{rightboundary}) will be explored in section~\ref{s:results}.
Before doing that, however, we explore the validity of the quasi-stationary approximation and two limiting cases that are particularly relevant for our purposes. 
In the first case, we verify that, as expected, our system reduces to the corresponding system describing $\ell$-boson stars when $s = i\omega$ is purely imaginary and one transforms the field $\psi_\ell$ to a Schwarzschild-type gauge. 
In the second case, we show that when the scalar field's amplitude converges to zero, one recovers the equations relevant for the scalar field configurations in the test field limit.

\subsection{Accretion time and validity of the quasi-stationary approximation} 
\label{s:AccretionRate}

Here, we briefly discuss the range of validity of the quasi-stationary approximation by analyzing the time evolution of the metric coefficients.
So far, we have not considered the equation for $\dot{m}$.
Using equations~(\ref{Eq:D0mEF},~\ref{Eq:mprimeEF}) one finds
\begin{equation}
\dot{m} = \frac{\kappa r^2}{2}\frac{a}{\gamma^2}\left\{
 \frac{2m}{r}\left[ T(n,n) + T(w,w) \right]
  + \left[ 1 + \left( \frac{2m}{r} \right)^2 \right] T(n,w) \right\},
\end{equation}
in the generalized Eddington-Finkelstein gauge. Next, using equations~(\ref{Eq:theta},~\ref{Eq:Tnn},~\ref{Eq:Tww},~\ref{Eq:Tnw}) one obtains
\begin{equation}
\dot{m} = \kappa_\ell r^2 a e^{2\sigma t}\left[ \frac{2m}{r}\left| \frac{s}{a}\psi_\ell \right|^2
 + \left( 1 - \frac{2m}{r} \right)\re\left( \frac{s^*}{a}\psi_\ell^*\psi_\ell' \right) \right].
\label{Eq:mdot}
\end{equation}
At the horizon $r = 2m = 2M$ and at $t=0$ this simplifies to
\begin{equation}
\dot{m_0} := \left. \dot{m} \right|_{t=0,r=2M} = \kappa_\ell\frac{ (2M)^2 |s|^2 |A|^2}{\left. a \right|_{t=0,r=2M}} ,
\end{equation}
which gives the mass accretion rate. Accordingly, we can introduce the time scale
\begin{equation} 
t_{a} := \frac{M}{\dot{m_0}} = \frac{\left. a \right|_{t=0,r=2M}}{4\kappa_\ell M|s|^2 |A|^2},
\label{taccr}
\end{equation}
and the quasi-stationary approximation is expected to be valid as long as
\begin{equation}
0\leq t \ll t_0:= \frac{1}{|\sigma|},\qquad
0\leq t \ll t_{a}.
\end{equation}

Additionally, we will find it useful to evaluate the quotients $t_0/t_L$
and $t_a/t_L$, which provide an estimate of the lifetime $t_0$ and the accretion time $t_a$ relative to the light-crossing time of the object, $t_L=2R_{99}$.
Here, we have defined, as usual, $R_{99}$ as the radius of a sphere containing 99\% of the object's mass.
As a reference, the universe's age relative to the Milky Way galaxy's light-crossing time is approximately 68,000. We will see that many of our solutions have $t_0/t_L$ and $t_a/t_L$ much larger than this value. 

\subsection{\texorpdfstring{$\ell$}{l}-Boson star limit}\label{sec.limit.boson.stars}

Next, we show that the eigenvalue problem~\eqref{Eq:eqns} for $(\psi_\ell,m,a)$ reduces to the corresponding problem determining $\ell$-boson stars when $s = i\omega$ is purely imaginary and the horizon boundary conditions~\eqref{boundary_conditions} are replaced with regularity conditions at the center.

To this purpose, one first needs to understand the relation between the generalized Eddington-Finkelstein coordinates $(t,r)$ and the Schwarzschild-type coordinates $(T,r)$, say, which are usually used to construct the boson star solutions. If the metric functions $a$ and $m$ depend on $r$ only, it is simple to verify that the coordinate transformation
\begin{equation}
t = T + \Theta(r),\qquad
\Theta(r) := \int^r \frac{2m(\tilde{r})}{\tilde{r} - 2m(\tilde{r})} \frac{d\tilde{r}}{a(\tilde{r})},
\end{equation}
brings the metric~(\ref{Eq:MetricEFGauge}) into the form
\begin{equation}
ds^2 = -a(r)^2\left( 1 - \frac{2m(r)}{r} \right) dT^2 + \frac{dr^2}{1 - \frac{2m(r)}{r}} + r^2 d\Omega^2.
\label{Eq:MetricSchwGauge}
\end{equation}
Of course, the metric functions $a(r)$ and $m(r)$ are unaffected by this transformation, since they only depend on $r$ by assumption. In contrast to this, the scalar field profile $\psi_\ell(r)$ which is related to the time-dependent scalar field $\phi_\ell(t,r)$ via the mode ansatz~\eqref{Eq:ModeAnsatz} is affected by this transformation since
\begin{equation}
\phi_\ell(t,r) = e^{st}\psi_\ell(r) = e^{sT} e^{s\Theta(r)}\psi_\ell(r),
\end{equation}
such that with respect to Schwarzschild-type coordinates, the scalar field profile is
\begin{equation}
\tilde{\psi}_\ell(r) = e^{s\Theta(r)}\psi_\ell(r).
\label{Eq:psitilde}
\end{equation}
If $s = i\omega$ is purely imaginary, this is only a phase difference; however, the derivative of the profile transforms non-trivially:
\begin{equation}
\tilde{\psi}'_\ell(r) 
 = e^{s\Theta(r)}\left[ \psi'_\ell(r) + \frac{s}{a}\frac{2m}{r - 2m}\psi_\ell(r) \right],
\end{equation}
and we see that the second term diverges at $r = 2m$.

Re-expressing the system~\eqref{Eq:eqns}
in terms of $\tilde{\psi}_\ell$ one finds:
\begin{subequations}
\label{Eq:Schw}
\begin{align}
\left(1 - \frac{2m}{r} \right)\tilde{\psi}_\ell'' 
 &= \tilde{B}\tilde{\psi}_\ell' + \tilde{C}\tilde{\psi}_\ell,
\label{Eq:PsiSchw}\\
2m' &= \kappa_\ell r^2\left[ \frac{1}{1 - \frac{2m}{r}}\frac{\omega^2}{a^2} \tilde{\psi}_\ell^2
 + \mu_\ell^2 \tilde{\psi}_\ell^2 + \left(1 - \frac{2m}{r} \right) \tilde{\psi}_\ell'^2 \right],
\label{Eq:mSchw}\\
\frac{a'}{a} &= \kappa_\ell r \left[ \tilde{\psi}_\ell'^2 
 + \frac{1}{1 - \frac{2m}{r}}\frac{\omega^2}{a^2} \tilde{\psi}_\ell^2 \right],
\label{Eq:aSchw}
\end{align}
\end{subequations}
where we have assumed without loss of generality that $\tilde{\psi}_\ell$ is real-valued, and where
\begin{align}
\begin{split}
\tilde{B} &:= B + \frac{4m}{r}\frac{s}{a},\\
\tilde{C} &:= C - \frac{s}{a}\frac{B}{1 - \frac{2m}{r}} + s\left( 1 - \frac{2m}{r} \right)
\left[ \frac{1}{a}\frac{1}{1 - \frac{2m}{r}} \right]'.
\end{split}
\end{align}
Taking into account that $\alpha := a\sqrt{1 - 2m/r}$ is the lapse, equations~(\ref{Eq:mSchw}, \ref{Eq:aSchw}) agree with equations~(7a) and (7b) in Ref.~\cite{Alcubierre:2018ahf} discussing the $\ell$-boson star configurations (note that in that reference $\gamma$ is defined as $\gamma = (1-2m/r)^{-1/2}$ which is different than in this article). 
Note also that equation~(\ref{Eq:mdot}) for $\dot{m}$ is transformed into
\begin{equation} 
\dot{m} = \kappa_\ell r^2\left( 1- \frac{2m}{r} \right)\re\left( s^*\tilde{\psi}_\ell^*\tilde{\psi}_\ell' \right),
\end{equation}
under the transformation~(\ref{Eq:psitilde}), whose right-hand side is zero when $s$ is purely imaginary and $\tilde{\psi}_\ell$ is real.

Although the transformation~(\ref{Eq:psitilde}) maps the problem~\eqref{Eq:eqns} for $(\psi_\ell,m,a)$ to the problem~\eqref{Eq:Schw} for $(\tilde{\psi}_\ell,m,a)$ which, together with appropriate regularity conditions at $r=0$ and asymptotic conditions at $r\to \infty$ yields the $\ell$-boson star solutions, it is by no means evident that the solutions of~\eqref{Eq:eqns} with the horizon boundary conditions~\eqref{boundary_conditions} should converge to regular solutions of~\eqref{Eq:Schw} in the limit $r_0 = 2M\to 0$. Therefore, this problem will be analyzed numerically in the next section.

\subsection{Test field limit}
\label{sec.test.field}

In the test-field limit, where the mass is constant 
$m(r)=M$ and  $a(r)=1$, as discussed at the beginning of Section~\ref{s:BC}, we introduce for convenience the scaled function 
$\frac{1}{r}\tilde u_{\ell} =\tilde \psi_{\ell}$, so that in this case the equation~\eqref{Eq:PsiSchw} can be written in terms of the variable $z$ as \cite{Barranco:2013rua}:
\begin{equation}
\label{eq:kg-testz}
-\left(   \frac{z}{z+1} \frac{d}{dz}     \right)^2\tilde u_{\ell}+
\left[   \frac{\ell(\ell+1)z}{(z+2)^3} + \frac{z}{(z+1)^4}-\frac{4(\mu M)^2}{z+1}                \right]\tilde u_{\ell}
= -\Omega^2 \tilde u_{\ell}\; ,
\end{equation}
where 
$\Omega=2M\sqrt{\mu^2+s^2}$.
Equation \eqref{eq:kg-testz} has two regular points at $z=0$ and $z=-1$, and an irregular point at $z=\infty$, and thus its solution can be written taking this into account, close to $z=0$.
Regular solutions that also describe the ingoing modes toward the horizon close to $z=0$ behave as $z^{-2Ms}$. Furthermore, close to $z=1$ the solution behaves as $(z+1)$.
In the asymptotic region $z\rightarrow \infty$, we are interested in solutions with exponential decay of the form $e^{-\Omega z}$.
A solution to equation \eqref{eq:kg-testz} that satisfies the required behavior near $z=0$ can be written in terms of the confluent Heun function $H_C$~\cite{heun} as 
\begin{equation}
\tilde u_{\ell}(z) = (z+1)z^{2Ms}e^{-\Omega z} H_C(\alpha,\beta,\gamma,1,2\Omega,-z) \; ,
\end{equation}
where 
\begin{align}
\alpha &= \ell(\ell+1) + 4M^2(\mu^2+2s^2)+\Omega(1+4Ms)-2Ms \; , \nonumber\\
\beta &=  4M^2(\mu^2+2s^2) + 2\Omega(1+2Ms) \; ,
\qquad
\gamma = 1+4Ms \; .
\nonumber
\end{align}
The function $H_C$ is defined with the convention used in \textsf{Mathematica}~\cite{math}. 
Furthermore, equation \eqref{eq:kg-testz} can be written  as a Schr\"odinger-like equation,
\begin{equation}   \label{KGtest}
    \left[ -\frac{\partial^2}{\partial {r^*}^2} + V_{\rm eff}(r^*)\right] \tilde{u}_\ell(r^*) = -s^2 \tilde{u}_\ell(r^*),
\end{equation}
with the effective potential given by
\begin{equation}
    V_{\rm eff}(r^*) := \left(1-\frac{2 M}{r}\right) \left[ \frac{\ell(\ell+1)}{r^2} + \frac{2 M}{r^3} + \mu^2\right], \qquad r=r(r^*),
\end{equation}
where the Regge-Wheeler tortoise coordinate $r^*:=r+2 M \ln(r/2M-1)\, \in (-\infty, \infty)$ has been used.

The effective potential has a local minimum if the following condition is satisfied ~\cite{Barranco:2011eyw}:
\begin{equation}
    (\mu M)^2 < -\frac{1}{32}(\ell^2+\ell-1)(\ell^2+\ell+1)^2 +\frac{1}{288}\sqrt{3(3\ell^4+6\ell^3+5\ell^2+2\ell+3)^3}\, ,
    \label{eq:mulimit}
\end{equation}
which is an indication that quasi-bound solutions for the scalar field can exist (see also table~\ref{t:cl}).

The test field values of $s=\omega + i \sigma$ in an approximation valid for small $\mu M$ are given in~\cite{Detweiler:1980uk}. The expressions are
\begin{subequations}
\begin{align}
    \frac{\omega_{n \ell}}{\mu} &\approx \sqrt{1 - \frac{(\mu M)^2}{(\ell+n)^2}}\, ,
    \label{testomega} \\
    \frac{\sigma_{n \ell}}{\mu} &\approx - \frac{ 2^{4\ell+4}\, (\mu M)^{4\ell+5}\, (2\ell+n)!\, \ell!^2\, \prod\limits_{j=1}^\ell \left(16\left(\mu M\right)^2+j^2\right)}{ (\ell+n)^{2\ell+4}\, (n-1)!\, (2\ell)!^2\, (2\ell+1)!^2 }\, .
    \label{testsigma}
\end{align}
\end{subequations}
Note that here we define $n$ such that $n=1$ gives the ground-state solution. Hence, the results shown here correspond to evaluating these expressions at $n=1$.
We note that the procedure in~\cite{Detweiler:1980uk} is in principle only valid for $\ell>0$. However, as we have verified in~\cite{Barranco:2012qs}, these expressions are also true for $\ell=0$.

\begin{figure}[tbh]
    \centering
    \includegraphics[width=0.45\linewidth]{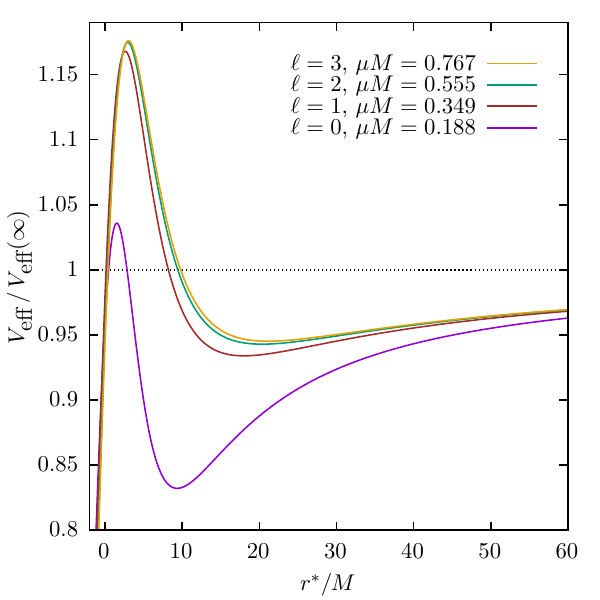}
    \includegraphics[width=0.44\linewidth]{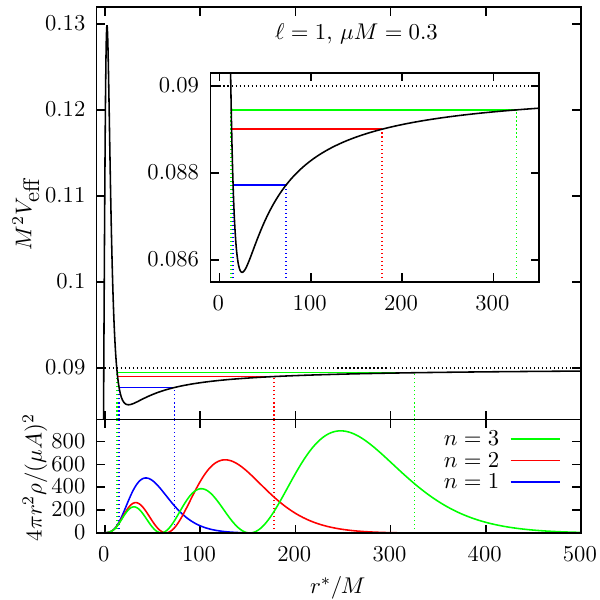}
    \caption{
    {\em Left panel:} Effective potentials with $\ell$ from $0$ to $3$. In each case, the value of $\mu M$ is chosen to be 75\% of the maximum for which there is still a well, as reported in table~\ref{t:cl}. 
    The potentials have been rescaled by their asymptotic value, $V_{\rm eff}(\infty)=\mu^2$.
    {\em Right panel:} Effective potential with ``energy levels'' $\omega_{n \ell}^2$ 
    and density profiles for $\ell=1$ and $\mu M=0.3$. 
    The vertical dotted lines indicate the ``classical turning points''.  
    }
    \label{f:veff}
\end{figure}

In figure~\ref{f:veff}, we show some effective potentials and some ``energy levels'' $\omega_{n \ell}$ and density profiles for an illustrative example with $\ell=1$ and $\mu M=0.3$. For the solutions shown in the figure, we have $|\sigma|\ll \omega$, or $s\approx\omega$.
We see that the energy levels accumulate at the asymptotic value of $V_{\rm eff}$, given in this case by $M^2V_{\rm eff}\to(\mu M )^2=0.09$. 
The density profiles show an oscillatory behavior inside the potential well, where $V_{\rm eff}<\omega_n^2$, having $n$ local maxima, and decay rapidly when $V_{\rm eff}>\omega_n^2$.

One can easily show that the potential well gets larger for larger $\ell$.
For instance, the location of the local minimum of $V_{\rm eff}$ at fixed $\mu M$ goes as $r_m\sim \ell^2$ for large $\ell$.
Furthermore, if one assumes the expression~\eqref{testomega} holds (which it usually does for $\mu M\lesssim0.1$), then it follows that the two classical turning points also scale as $r_l\sim r_r\sim \ell^2$. Hence, the cases with larger values of $\ell$ admit solutions with larger radii and with a larger ``hollow'' central region.

Although this analysis, in terms of an effective potential, is only valid for the test-field approximation, we will see that qualitatively similar results are found in in the self-gravitating case.

\FloatBarrier
\section{Noble gravitational atom solutions} \label{s:results}
To obtain noble gravitational atom solutions, we integrate numerically equations~\eqref{Eq:eqns} with $e^{2\sigma t}=1$,
imposing boundary conditions at $r_0$  and $r_1$ as in equations~\eqref{boundary_conditions} and~\eqref{rightboundary}, using a shooting algorithm. 
The numerical code is based on~\cite{Press1996Numerical}, but with the adaptive-step solver from~\cite{Radhakrishnan1993}, which provides high accuracy at low computational cost. The code was first presented and tested in~\cite{Megevand:2007uy}. 

Since we prefer to express the solutions so that $a(r_1)=1$, instead of $a(r_0)=1$, we rescale them as $a(r)\mapsto a(r)/a(r_1)$, $s\mapsto s/a(r_1)$.
In this way, we recover the Schwarzschild metric at large $r$, with $m(r\to\infty)=M_T$ the total mass of the configuration, including both the black hole and the surrounding scalar field distribution. 
Furthermore, it is easy to see that the solutions have the following rescaling properties,
\begin{equation}
(\mu, s) \mapsto \lambda (\mu, s),\; (r, m)  \mapsto \lambda^{-1} (r, m),\, (a, \psi_\ell) \mapsto (a, \psi_\ell), 
\end{equation}
which allow us to represent them in terms of just three parameters: $\ell$, $\mu M$, $A$.
In this section, we concentrate on solutions with $\ell=0$, $1$, and $2$, since they show distinctive features, whereas for $\ell\ge2$ the solutions are qualitatively very similar for different values of $\ell$. 
For reference, we include some results with $\ell=3$ in appendix~\ref{app.ell=3}. 
Finally, we will mostly restrict ourselves to the nodeless ``ground-sate'' solutions, for which $n=1$.

Although it is not easy to demonstrate an analogous condition to that shown in equation~\eqref{eq:mulimit} in the self-gravitating case, since such an effective potential is not clearly defined, we note that we were able to find solutions with $\mu M$ close, but not beyond the test field expression. 
For convenience, we evaluate this limit for the first few values of $\ell$ and show them in table~\ref{t:cl}.

\begin{table}[htb]
    \centering
    \caption{For the first few values of $\ell$, we show the maximum value of $\mu M$ for which test field quasi-stationary solutions can exist, given in equation~\eqref{eq:mulimit}, and the factor $c_\ell$, defined in equation~\eqref{eq:cl}.}
    \begin{tabular}{ccc}
    \hline
    \hline
       $\ell$ & $\mu M$ limit & $c_\ell$\\
       \hline
       0  & 0.25 & 1 \\
       1  & 0.4656 & 0.5048  \\
       2  & 0.7401 & 0.174   \\
       3  & 1.023 & 0.00366  \\   
    \hline
    \hline
    \end{tabular}
    \label{t:cl}
\end{table}

Previously, in Ref.~\cite{Alcubierre:2024mtq}, we found that, in a certain limit, an $\ell=0$ self-gravitating gravitational atom is practically indistinguishable from an $\ell=0$ boson star of the same central amplitude, except at the very center.
For $\ell> 0$, identifying the appropriate $\ell$-boson star to compare with a given noble gravitational atom is less obvious.
To obtain $\ell$-boson star solutions, one solves the Einstein-Klein-Gordon equations in terms of $u(r)$, which is defined from $\psi(r)$ according to
\begin{equation}
    \psi(r)=r^\ell u(r)
\end{equation}
(see~\cite{Alcubierre:2018ahf}).
Evaluating this equation at the event horizon, 
using $u(r)= u_0+\mathcal{O}(r^2)$ for the $\ell$-boson star, and reading the value $\psi(2M)=A$ for the noble gravitational atom, we obtain, 
to first order, 
\begin{equation}
    A \approx (2 M)^\ell u_0. \label{eq:Au0}
\end{equation}
Hence, to properly compare noble gravitational atoms with $\ell$-boson stars, one might expect that the scalar field amplitudes to choose for each pair of objects should satisfy equation~\eqref{eq:Au0}, with $A$ for the noble gravitational atom and $u_0$ for the $\ell$-boson star.
Note that this approximation works well for $\ell=0$ (in which case it simply reduces to $A=u_0$), although it is not completely clear why it should. In fact, we will see that it fails for $\ell>0$, obtaining instead
\begin{equation}
    A = c_\ell (2 M)^\ell u_0 \label{eq:cl},
\end{equation}
where $c_\ell$ is found empirically to be given by the values shown in table~\ref{t:cl}.
Finally, it is convenient to define $\bar{u}_0:=u_0/\mu^\ell$, and then write
\begin{equation}
A=c_\ell (2\mu M)^\ell \bar{u}_0,
\label{u0tilde}
\end{equation}
where $A$, $c_\ell$, $\mu M\equiv\alpha_G$ and $\bar{u}_0$ are all dimensionless.

\subsection{Density profiles of solutions with \texorpdfstring{$\ell=0$, $1$, and $2$}{l=0, 1, and 2}}
\subsubsection{\texorpdfstring{$\ell=0$}{l=0}:}

\begin{figure}[tbh] \centering
\includegraphics[width=0.66\textwidth]{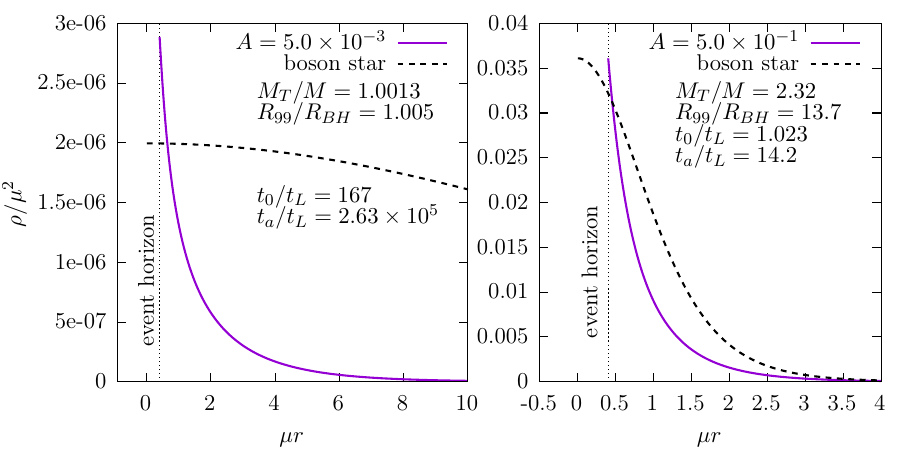}
    \caption{ \justifying
     Density $\rho$ of gravitational atoms with $\ell=0$, $\mu M=0.2$, and different amplitudes $A$. For large $\mu M$, the density profiles tend to be sharp and do not resemble a boson star for any value of $A$.
    \label{f:l0rho_comp1}} 
\end{figure}

\begin{figure}[tbh] \centering
\includegraphics[width=0.66\textwidth]{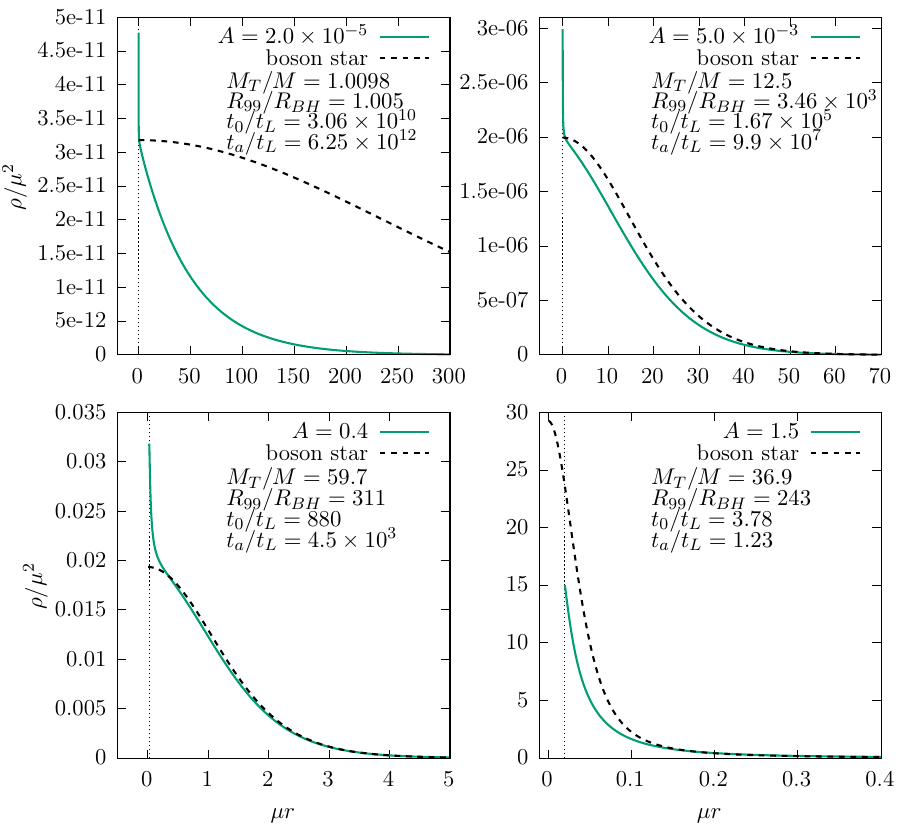}
    \caption{ \justifying
     Density $\rho$ of gravitational atoms with $\ell=0$, $\mu M=0.01$, and different amplitudes $A$. For intermediate $\mu M$, some configurations become wider, and for certain values of $A$, they start to resemble boson stars.
    \label{f:l0rho_comp2}} 
\end{figure}

\begin{figure}[tbh] \centering
\includegraphics[width=0.99\textwidth]{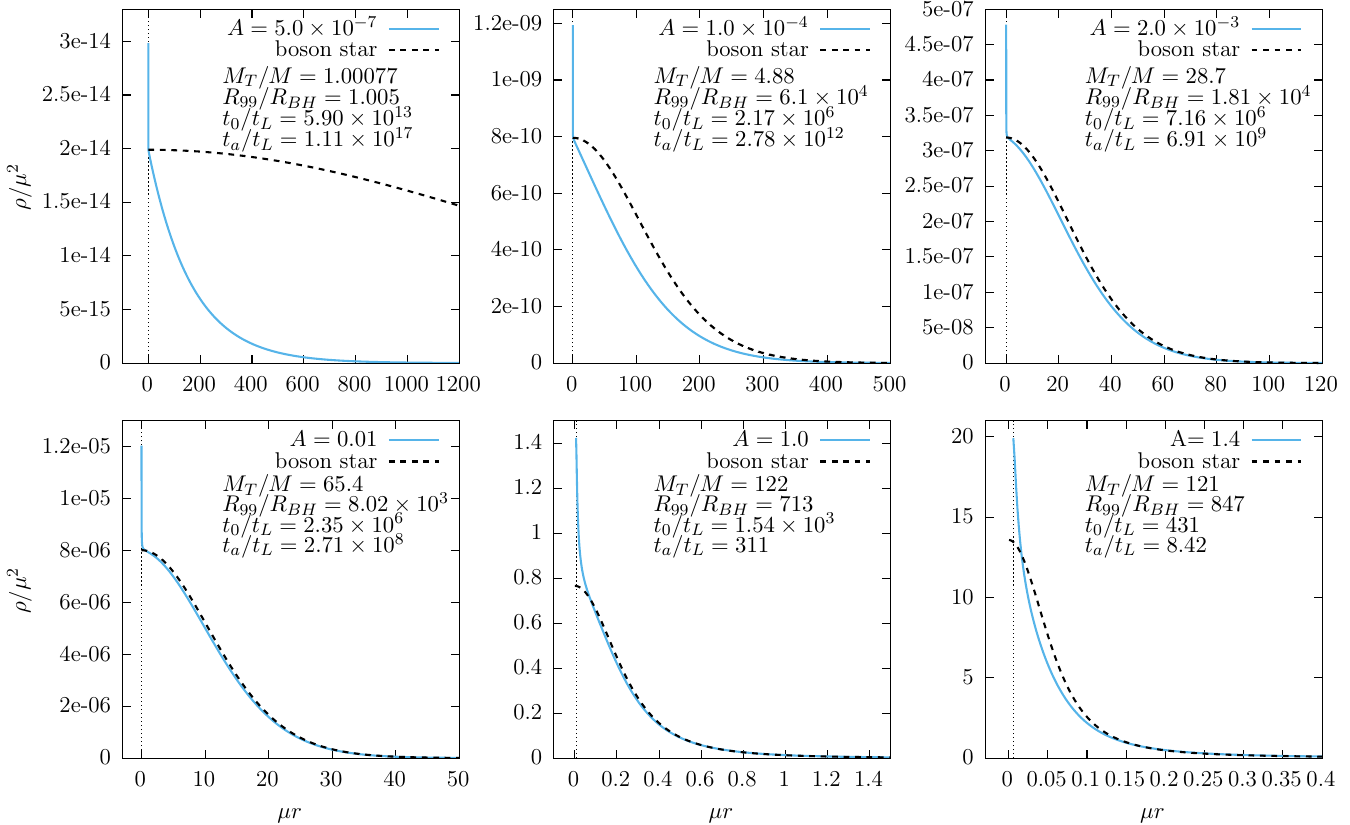}
    \caption{ \justifying
    Density $\rho$ of gravitational atoms with $\ell=0$, $\mu M=0.003$, and different amplitudes $A$. For small $\mu M$, some configurations are identical to boson stars, except for the black hole spike at their center. 
    \label{f:l0rho_comp3}} 
\end{figure}

We first consider the case $\ell=0$ (see also our previous work in~\cite{Barranco:2017aes, Alcubierre:2024mtq} for more details). 
In figures \ref{f:l0rho_comp1}, \ref{f:l0rho_comp2}, and \ref{f:l0rho_comp3} we show density plots with $\ell=0$ and $\mu M=0.2$, $0.01$, and $0.003$, respectively, for different values of $A$, which we have chosen to illustrate the distinct qualitative behaviors exhibited by the solutions. The vertical dotted line indicates the event horizon's location.

We can see important differences among all the solutions' properties. 
For large $\mu M$, close to the maximum limit $\mu M =0.25$ given by equation~\eqref{eq:mulimit}, the density profiles tend to be sharp for all values of $A$, concentrating mostly towards the event horizon, as can be clearly seen in figure~\ref{f:l0rho_comp1}. 
As $\mu M$ becomes smaller, figures~\ref{f:l0rho_comp2} and~\ref{f:l0rho_comp3}, we start to see a distinctive qualitative change for intermediate values of $A$. 
In this region, the solutions become wider and tend to become flat towards the center, except for a very sharp spike that begins to appear right at the event horizon. 
These profiles start to resemble those of boson stars, as is more evident from figure~\ref{f:l0rho_comp3}. 
In all these figures where we have included the density of the boson stars for reference.
For all values of $\mu M$, the scalar field configurations become less compact as $A$ becomes small, reaching the test field solutions for $A$ sufficiently small.\footnote{We distinguish here the scalar field main distribution from the whole object, since the object's compactness in that limit becomes maximal as the scalar field vanishes and we are left with just a black hole.} 
Towards the other end, the compactness increases with $A$ until the solution with maximum mass is reached.

We concentrate now on the comparison with boson stars. In the $\ell=0$ case, by simply setting the gravitational atom and boson star amplitudes to coincide, we can obtain a perfect match at large scale between these two solutions, provided that $\mu M$ is small enough and that the amplitudes $A$ lie in a certain intermediate range (see figures~\ref{f:l0rho_comp3} and~\ref{f:l0Mw}). 
Such solutions distinguish themselves from boson stars only at the very sharp spike that appears at the event horizon, yielding a good match even relatively close to the event horizon. 
This is most clearly seen in the fourth panel of figure~\ref{f:l0rho_comp3}.

For large values of $\mu M$ (figure~\ref{f:l0rho_comp1}), we do not find any values of $A$ for which the solutions resemble boson stars. This can also be noticed in figure~\ref{f:l0Mw}. From all the solutions found with $\mu M=0.2$, we only show in figure~\ref{f:l0rho_comp1} those with the smallest and largest $A$, since all intermediate cases look qualitatively similar to them, in contrast with what happens for smaller values of $\mu M$ (see, for instance, figure~\ref{f:l0rho_comp3}).
In the cases with $\mu M=0.01$ (figure~\ref{f:l0rho_comp2}), we see that, for intermediate values of $A$, the profile seems to start approaching one similar to that of a boson star, but $\mu M$ is still too large to give a good match. 

As we have seen before, in~\cite{Alcubierre:2024mtq}, and figure~\ref{f:l0rho_comp3} illustrates well, for a small enough $\mu M$ ($=0.003$ in this case), there is a range of the amplitude $A$ for which the solutions resemble boson stars at large scale, away from the black hole, and display a small ``spike" near the event horizon. 
    For smaller values of $A$, the scalar field's self-gravity becomes negligible, so we can no longer have something resembling a boson star. In this limit, we recover the test field gravitational atoms obtained in~\cite{Barranco:2011eyw, Barranco:2012qs}, as expected.  
    As $A$ becomes larger, the region resembling a boson star becomes more compact, and the spike becomes more prominent, reaching a point where we again lose the boson star resemblance completely.
    In the second-to-last plot, we can still distinguish a spike (note the abrupt slope change close to the event horizon), but this distinction is completely gone in the last plot. 
    We have not been able to identify the physical mechanism responsible for the change in behavior that appears beyond a certain value of $A$, but it seems to be characteristic of all values of $\ell$, as we will also see later. 
    In all cases where a spike is present, we find its contribution to the mass to be negligible.

We have included in the figures the values of some quotients to give an idea of the relative mass, size, lifetime, and accretion time of each object. They are, respectively, $M_T/M$, $R_{99}/R_{BH}$, $t_0/t_L$, and $t_a/t_L$, where $R_{BH}=2 M$. 
Taking, for instance, the values shown in the fourth plot of figure~\ref{f:l0rho_comp3}, which exhibits a closer similarity between the gravitational atom and the boson star, we see that some of these objects can be very large, very dilute, and very long-lived. 
In the next subsections, we will see that these characteristics are not uncommon for those noble gravitational atoms that resemble $\ell$-boson stars.

Before proceeding with the description of solutions with $\ell>0$, we address the spikes present in the $\ell=0$ case. We do so by comparing, in figure~\ref{f:rhospike}, one of our solutions with one given by a cold dark matter model, as presented in appendix~\ref{app.CDM}. 
For our model, we choose a solution already considered in~\cite{Alcubierre:2024mtq} to describe the dark matter core of the Milky Way.
Going back for a moment to physical units, we have that its parameters are $\mu c^2=10^{-22}{\rm eV}$, $M=4.3\times 10^6 M_\odot$ (so that $\alpha_G=3.2\times 10^{-6}$), and $A=1.7\times 10^{-7}$. 
The characteristic times of this solution are $t_0\gtrsim 10^{11} {\rm years}$ and $t_{a}=3\times 10^{15} {\rm years}$. 
Its mass and radius are $M_T=10^9 M_\odot$ and $R_{99}=750 {\rm pc}$.

Note that the two spikes differ significantly in shape and amplitude, with the cold dark matter one being both steeper and more extended. This can be seen more clearly in the inner panel of figure~\ref{f:rhospike}.
To quantify the differences, we use the phenomenological profile $\rho(r) \propto (R_{\rm{sp}}/r)^\gamma$, which has been widely used in the literature to describe black hole spikes.
The slope $\gamma$ typically lies in the range $1.5 < \gamma < 2.5$, depending on the physical assumptions regarding the dark matter model and the astrophysical environment~\cite{Gondolo:1999ef, Ullio:2001fb}. The spike radius $R_{\rm{sp}}$ is more uncertain, but it is expected to be of the order of a few parsecs to a few tens of parsecs~\cite{Lacroix:2018zmg}.
Fitting our numerical solution to this profile results in an extremely shallow spike, with $\gamma\sim 0.15$, together with an unusually small characteristic radius $R_{\rm{sp}}\sim 10^{-5}\,\textrm{pc}$. 
Such values lie well outside the standard expectations for collisionless cold dark matter, suggesting that in the scalar field model other processes,  such as the gradient pressure associated with the ultralight dark matter mass, may prevent the formation of a sharp cusp.

\begin{figure}[tbh] \centering
\includegraphics[width=0.5\textwidth]{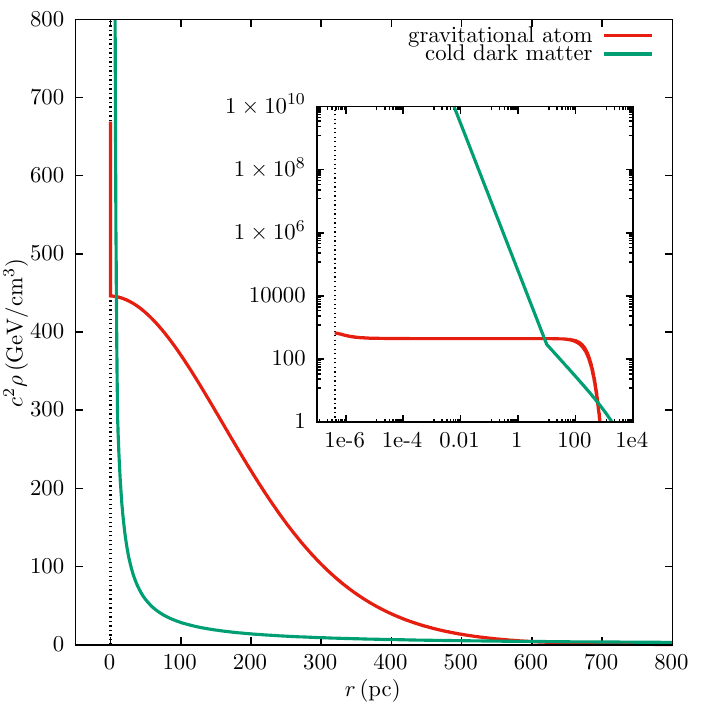}
    \caption{ \justifying
    Spike comparison between an $\ell=0$ gravitational atom (red line) and a cold dark matter model (green line), both constructed with parameters appropriate for the Milky Way. 
    \label{f:rhospike}} 
\end{figure}

\FloatBarrier
\subsubsection{\texorpdfstring{$\ell=1$}{l=1}:}

\begin{figure}[tbh] \centering
\includegraphics[width=0.66\textwidth]{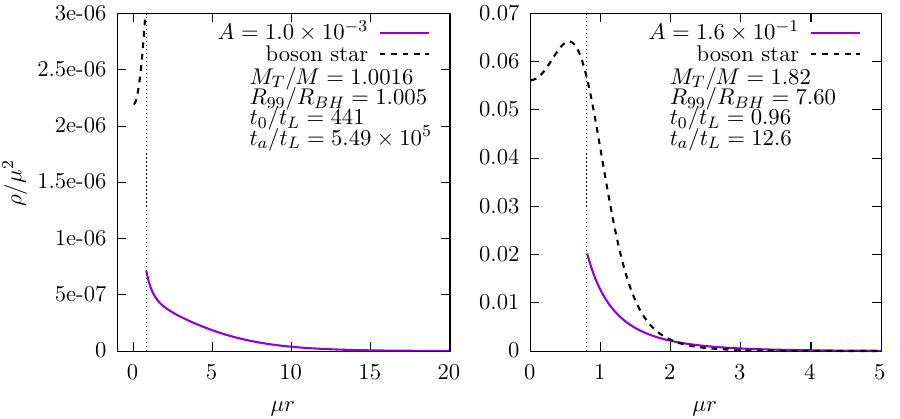}
    \caption{ \justifying
    Density $\rho$ of noble gravitational atoms with $\ell=1$, $\mu M=0.4$, and different amplitudes $A$.  As for the $\ell=0$ case, for large $\mu M$ the density profiles tend to be sharp and do not resemble a $\ell=1$ boson star for any value of $A$. 
    \label{f:l1rho_comp0}} 
\end{figure}

\begin{figure}[tbh] \centering
\includegraphics[width=0.99\textwidth]{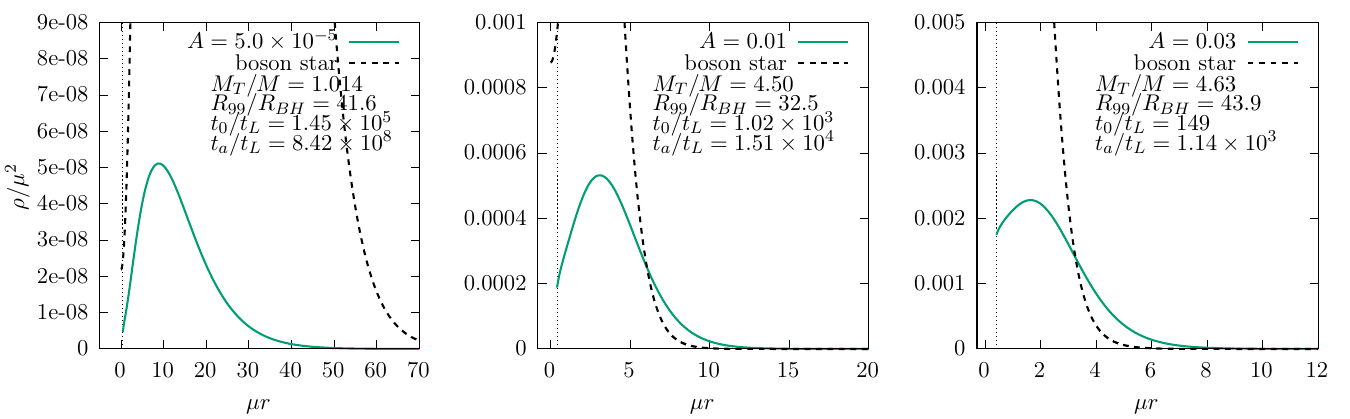}
    \caption{ \justifying
    Density $\rho$ of noble gravitational atoms with $\ell=1$, $\mu M=0.2$, and different amplitudes $A$. 
    \label{f:l1rho_comp1}} 
\end{figure}

\begin{figure}[tbh] \centering
\includegraphics[width=0.66\textwidth]{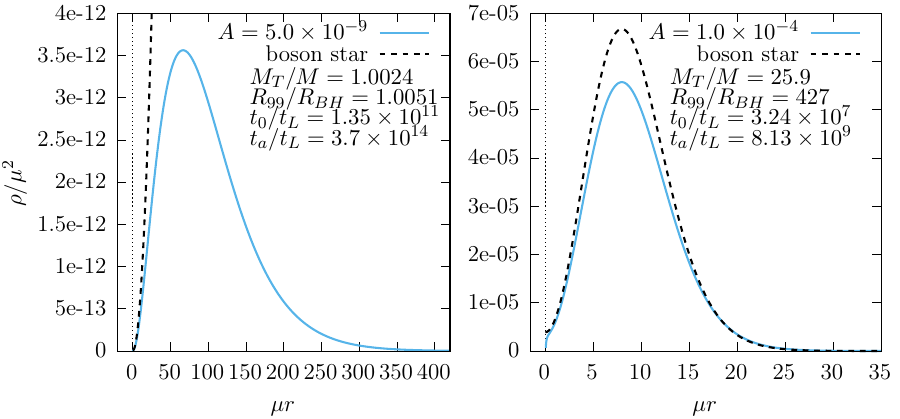}
    \caption{ \justifying
    Density $\rho$ of noble gravitational atoms with $\ell=1$, $\mu M=0.03$, and different amplitudes $A$.
    \label{f:l1rho_comp2}} 
\end{figure}

In figures~\ref{f:l1rho_comp0}, \ref{f:l1rho_comp1} and \ref{f:l1rho_comp2}, we show density profiles for solutions with $\ell=1$. 
For sufficiently small values of $\mu M$, and for suitable choices of $A$,
we now find profiles for which their maxima are located away from the central region, decreasing towards the event horizon, where they reach some value greater than zero. These features coincide with those of $\ell$-boson stars with $\ell$ also equal to $1$.


\begin{figure}[tbh] \centering
\includegraphics[width=0.45\textwidth]{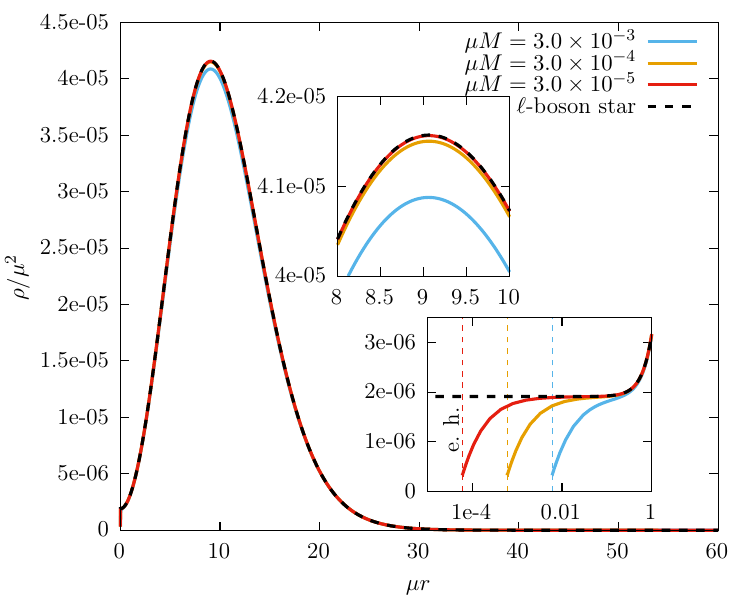}
\includegraphics[width=0.45\textwidth]{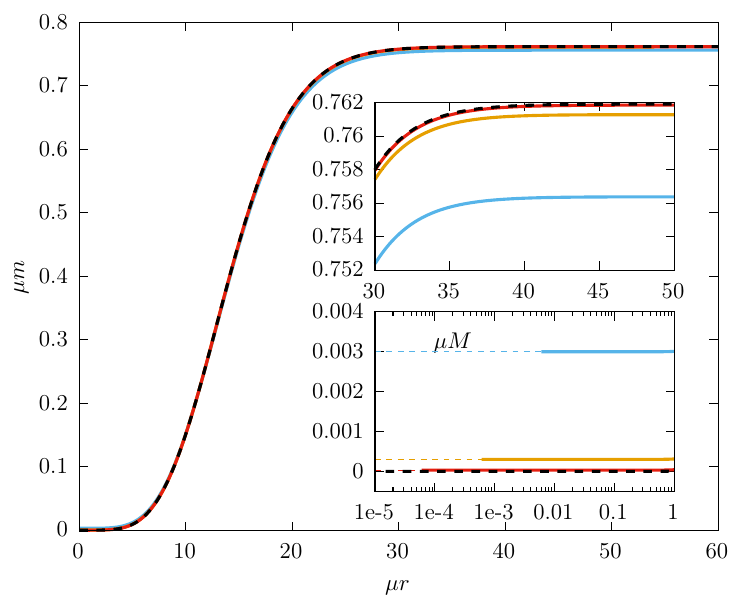}
    \caption{ \justifying
    Noble gravitational atom and $\ell$-boson star solutions with $\ell=1$ and amplitudes $\bar{u}_0=2.311\times 10^{-3}$ and $A$ such that their profiles coincide at small $\mu M$, according to equation~\eqref{u0tilde}. We plot the mass function $m(r)$ and the energy density $\rho(r)$ for three values of $\mu M$.  
    \label{f:l1comp2}} 
\end{figure}

Figure~\ref{f:l1comp2} shows density and mass plots of noble gravitational atoms and an $\ell$-boson star, all with $\ell=1$, with amplitudes $A$ and $u_0$ chosen so that their profiles coincide at large scale provided that $\mu M$ is small enough.
To obtain the proper limit in the case $\ell=1$, we find empirically that we need to compare solutions such that 
$A = 0.5048 \,(2M) u_0,$
as opposed to $A=2M u_0$, which would be implied by equation~\eqref{eq:Au0}. 
Contrary to the excellent matching seen at large scale for small enough $\mu M$, we observe differences in the region close to the black hole. 
In particular, close to the horizon we see a small dip, in contrast to the small spikes that appear when $\ell=0$~\cite{Alcubierre:2024mtq}.
Note, however, that those small differences in the density have a negligible effect on the mass profile.

\FloatBarrier
\subsubsection{\texorpdfstring{$\ell=2$}{l=2}:}

\begin{figure}[tbh] \centering
\includegraphics[width=0.66\textwidth]{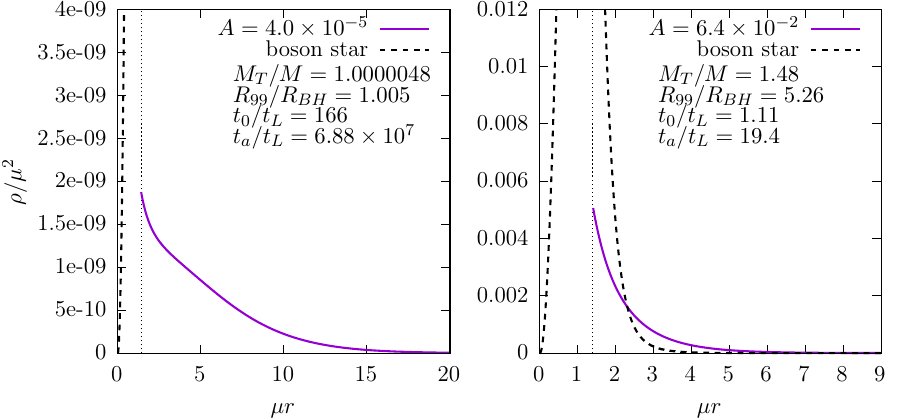}
    \caption{ \justifying
    Density $\rho$ of noble gravitational atoms with $\ell=2$, $\mu M=0.7$, and different amplitudes $A$. For large $\mu M$, the density profiles tend to be sharp and do not resemble an $\ell=2$ boson star for any value of $A$. 
    \label{f:l2rho_comp1}} 
\end{figure}

\begin{figure}[tbh] \centering
\includegraphics[width=0.99\textwidth]{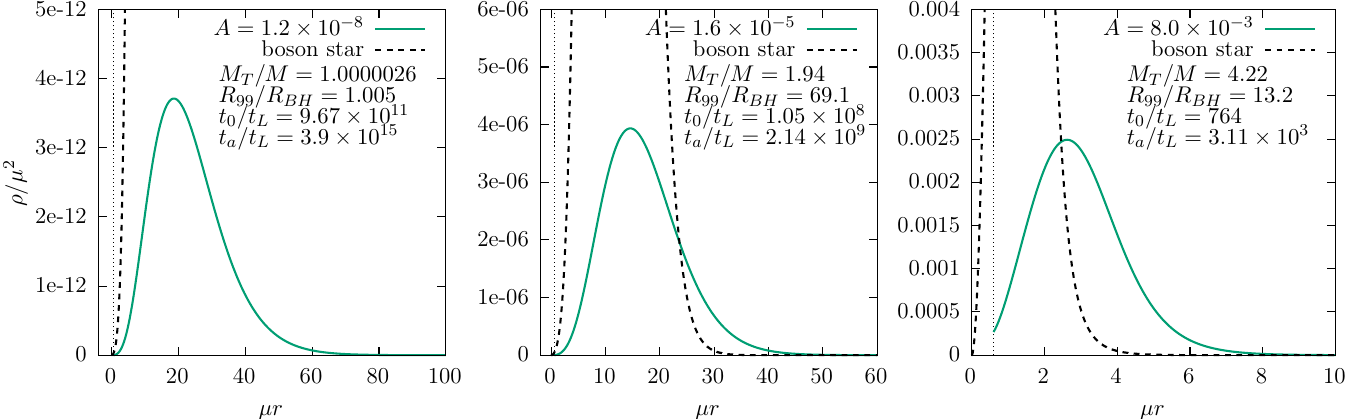}
    \caption{ \justifying
    Density $\rho$ of noble gravitational atoms with $\ell=2$, $\mu M=0.3$, and different amplitudes $A$. For small $\mu M$, some configurations are identical to $\ell=2$ boson stars, except for what happens at their center. 
    \label{f:l2rho_comp2}} 
\end{figure}

Finally, in figures~\ref{f:l2rho_comp1} and \ref{f:l2rho_comp2}, we show density profiles for solutions with $\ell=2$. As was the case for $\ell=1$, some profiles have a maximum away from the central region, with some of those decreasing towards zero at the horizon, except for a small spike or dip there in some cases. 
Once again, we find a resemblance with $\ell$-boson stars of the same $\ell$.
The similarities seem to occur for all values of $\ell$. See also the appendix, where we show results for $\ell=3$.

\begin{figure}[tbh] \centering
\includegraphics[width=0.45\textwidth]{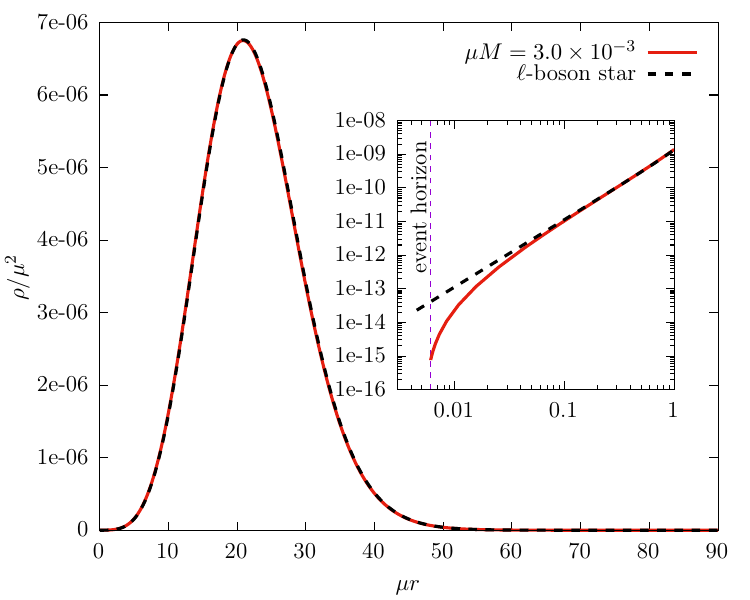}
\includegraphics[width=0.45\textwidth]{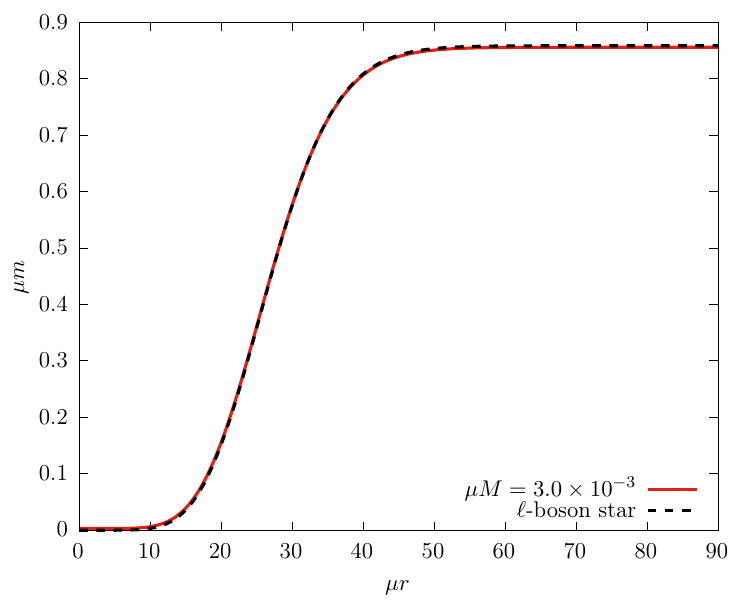}
    \caption{ \justifying
    Noble gravitational atom and $\ell$-boson star solutions with $\ell=2$, $\mu M=0.003$, and amplitudes $\bar{u}_0=2.39\times 10^{-5}$ and $A$ such that their profiles coincide at large scale. 
    \label{f:l2comp}} 
\end{figure}

In figure~\ref{f:l2comp}, we show the density profiles of a noble gravitational atom and a matching $\ell$-boson star, both with $\ell=2$. 
Since, as we know, $\rho(r)$ goes to $0$ at the center for $\ell$-boson stars with $\ell>1$, it was expected to see a better coincidence in this case (compare with the $\rho(r)$ plot in figure~\ref{f:l1comp2}). 
We find that the amplitude pairs that give matching profiles at large scales between noble gravitational atoms and $\ell$-boson stars  satisfy
$A = 0.174\, (2M)^2 u_0$,
so we obtain $c_2=0.174$.

\FloatBarrier
\subsection{Spectrum of solutions with \texorpdfstring{$\ell=0$, $1$, and $2$}{l=0, 1, and 2}}
We now continue the description of the solutions, focusing particularly on aspects related to their spectrum.

\subsubsection{\texorpdfstring{$\ell=0$}{l=0}:}
\begin{figure}[tbh] \centering
\includegraphics[width=0.49\textwidth]{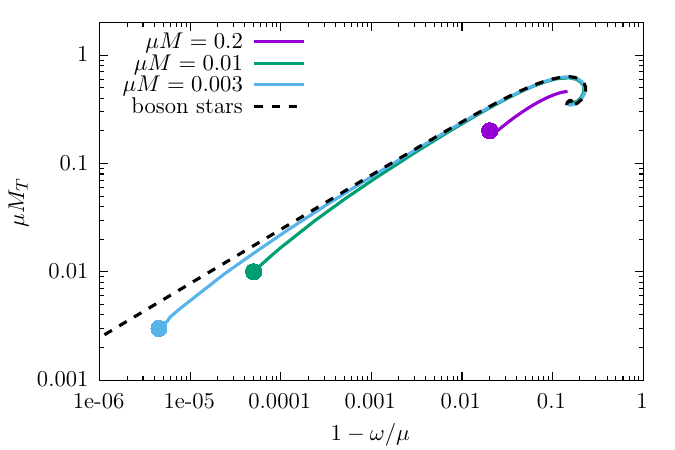}
    \caption{ \justifying
    $M_T$ vs $\omega$ of gravitational atoms and boson stars with $\ell=0$. The curve colo  rs match those used in figures~\ref{f:l0rho_comp1}, \ref{f:l0rho_comp2}, and~\ref{f:l0rho_comp3} for the given value of $\mu M$. 
    \label{f:l0Mw}} 
\end{figure}

Figure~\ref{f:l0Mw} shows a plot of the total mass $M_T$ vs frequency $\omega$ for varying values of the amplitude in the case $\ell=0$, so that each point in the curves represents a solution with a given $A$. Also shown, for comparison, is the corresponding curve describing $\ell=0$ boson stars.  
The gravitational atom curves have accumulation points toward the left as $A$ tends to $0$, where the test field limit is obtained.
Toward the other end, we have decided to stop the curves in each case either at some arbitrary point past the total mass maximum,\footnote{It is well known that $\ell$-boson star solutions located past that maximum are unstable~\cite{Alcubierre:2019qnh, Alcubierre:2021mvs}. Note, however, that a stability analysis for self-gravitating gravitational atoms has yet to be performed. We leave this study for future work.} or at the maximum $A$ for which we were able to find a solution.
For small enough $\mu M$, we see regions of $A$ values where the gravitational atom curves approach the boson star curve.
The thick dots indicate the test field limit in the small $\mu M$ approximation, with $M_T=M$ and $\omega$ given by equation~\eqref{testomega}. As expected, these dots are accumulation points for the solutions whenever $\mu M$ is small enough. Here, we see a small deviation only in the case of $\mu M=0.2$.

\begin{figure}[tbh] \centering
\includegraphics[width=0.99\textwidth]{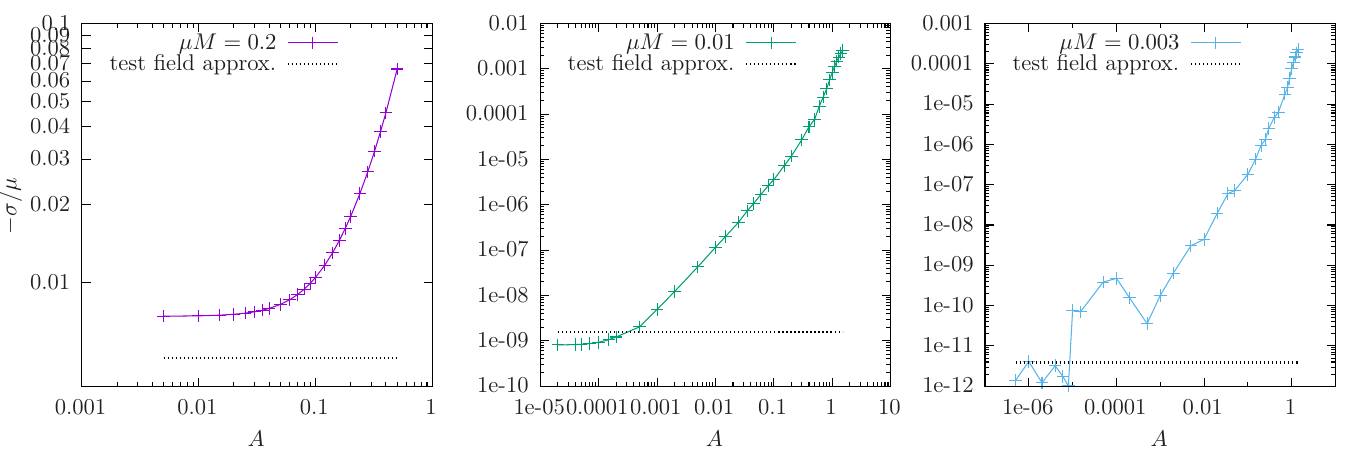}
    \caption{ \justifying
    $\sigma$ vs $A$ of gravitational atoms with $\ell=0$. The curve colors match those used in figures~\ref{f:l0rho_comp1}, \ref{f:l0rho_comp2}, and~\ref{f:l0rho_comp3} for the given value of $\mu M$.
    \label{f:l0sigma}} 
\end{figure}

Figure~\ref{f:l0sigma} shows plots of the frequency's imaginary part, $\sigma$, vs $A$ for different values of $\mu M$.
We observe that the curves have an asymptote, as expected, since $\sigma$ should converge to its test field value as $A$ becomes small. 
    We include, as a dashed line, the test field approximation of $\sigma$ valid at small values of $\mu M$, equation~\eqref{testsigma}. 
    Since $\mu M$ in the left plot is close to the maximum allowed value for the test field limit, this approximation is not so good. As $\mu M$ becomes smaller, the approximation seems to improve. 
    Unfortunately, we also obtain a curve with larger errors in that limit, making it difficult to corroborate the expected asymptotic value.
The relative errors in $\sigma$ start to grow as $\sigma$ itself becomes very small, making it in some cases numerically indistinguishable from zero. 
This is not uncommon when trying to evaluate a quantity that is very close to zero numerically.
However, we highlight that those errors in $\sigma$ never translate into large errors in any of the other quantities. In those regimes, we have $|\sigma|\ll \omega$ (with $\omega$ obtained with very small error), so that $s=\omega+i\sigma\approx\omega$ and the solutions become ``almost" stationary, with the errors in $s$, and hence also in the other quantities, remaining very small. 
For astrophysically relevant cases, these values of $\sigma$ correspond to decay times that already exceed the age of the Universe, even within the error. Hence, improving the numerical method to reduce these errors further may not be particularly meaningful.

\FloatBarrier
\subsubsection{\texorpdfstring{$\ell=1$}{l=1}:}
\begin{figure}[tbh] \centering
\includegraphics[width=0.5\textwidth]{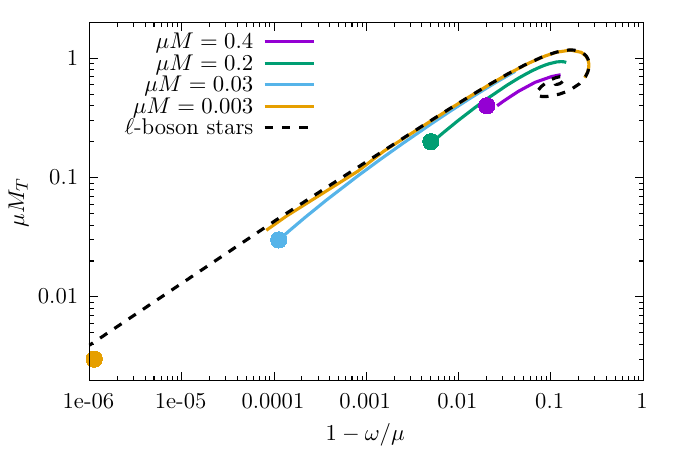}
    \caption{ \justifying
    $M_T$ vs $\omega$ of noble gravitational atoms and $\ell$-boson stars with $\ell=1$. 
    \label{f:l1Mw}} 
\end{figure}

\begin{figure}[tbh] \centering
\includegraphics[width=0.66\textwidth]{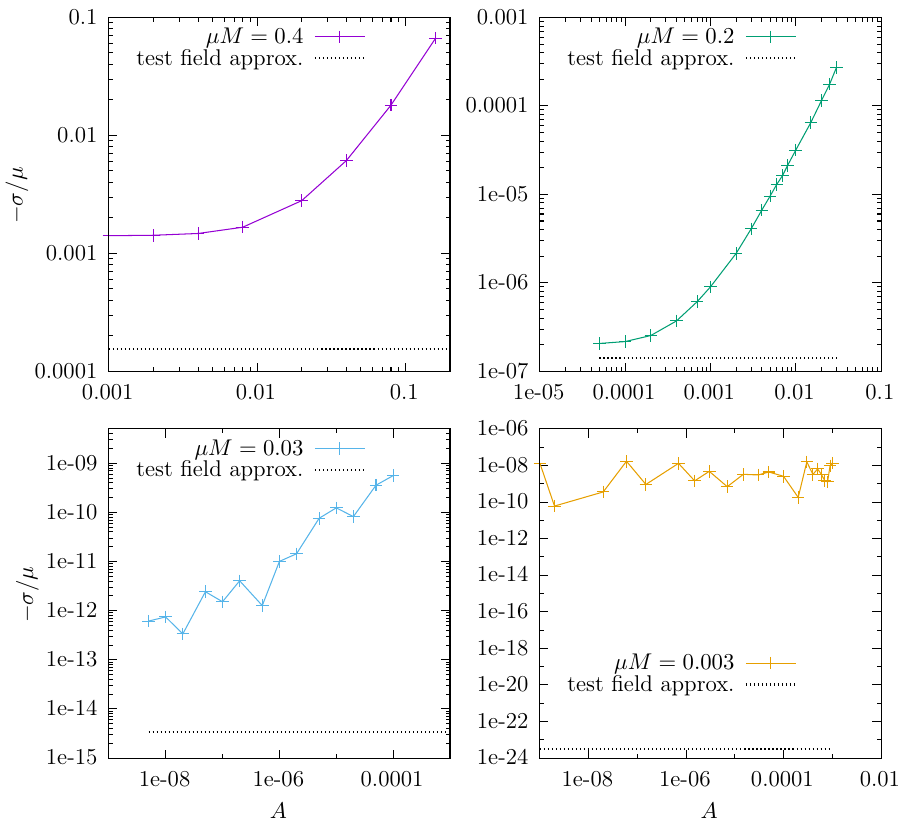}
    \caption{ \justifying
    $\sigma$ vs $A$ of noble gravitational atoms with $\ell=1$.  
    \label{f:l1sigma}} 
\end{figure}

We move on now to the case with $\ell=1$.
Some results are shown in figures~\ref{f:l1Mw} and~\ref{f:l1sigma}.
Figure~\ref{f:l1Mw} shows the total mass vs frequency for varying amplitude.
In this case, we were not able to obtain solutions near the test field limit in the case $\mu M=0.003$, so the curve ends before reaching the corresponding dot.
Figure~\ref{f:l1sigma} shows plots of the density's imaginary part $\sigma$ vs amplitude $A$ for different values of $\mu M$. As was the case for $\ell=0$, the relative errors in $\sigma$ become large for small $\mu M$, as $\sigma$ itself becomes very small.  

\FloatBarrier
\subsubsection{\texorpdfstring{$\ell=2$}{l=2}:}

\begin{figure}[tbh] \centering
\includegraphics[width=0.5\textwidth]{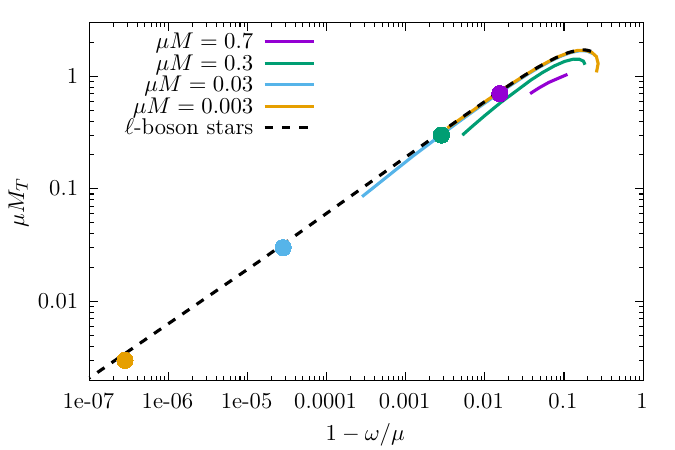}
    \caption{ \justifying
    $M_T$ vs $\omega$ of noble gravitational atoms and $\ell$-boson stars with $\ell=2$. 
    \label{f:l2Mw}} 
\end{figure}

\begin{figure}[tbh] \centering
\includegraphics[width=0.66\textwidth]{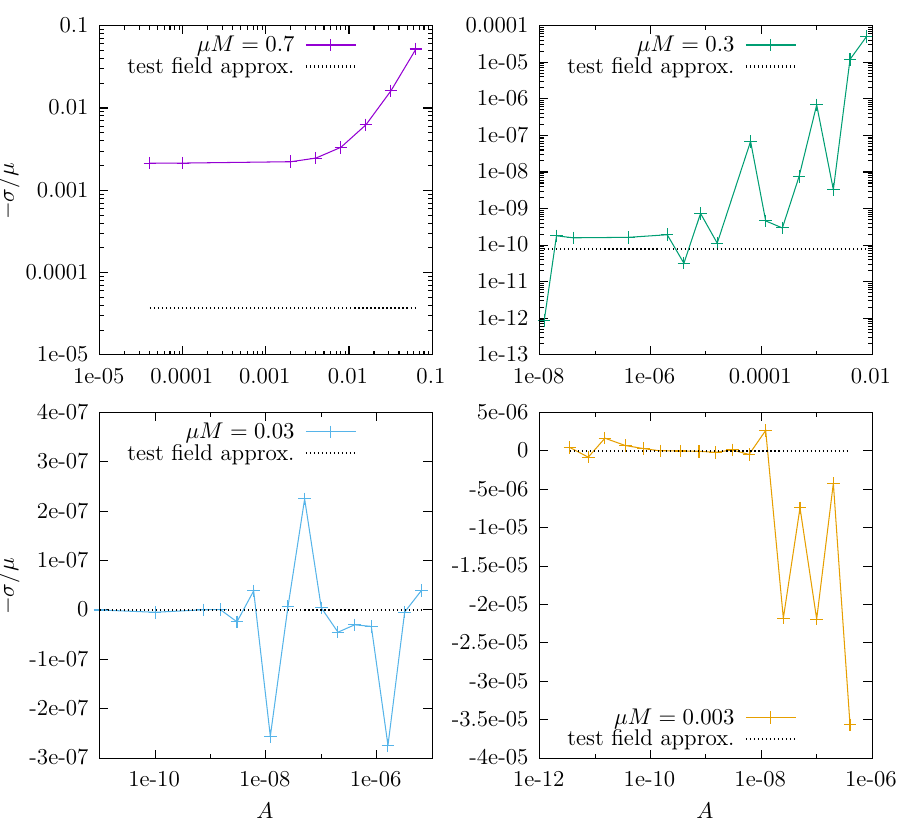}
    \caption{ \justifying
    $\sigma$ vs $A$ of noble gravitational atoms with $\ell=2$. 
    \label{f:l2sigma}} 
\end{figure}

For the case with $\ell=2$, we show some results in figures~\ref{f:l2Mw} and~\ref{f:l2sigma}.
Figure~\ref{f:l2Mw} shows a plot of the total mass vs frequency for varying amplitude, for $\ell$-boson stars and noble gravitational atoms with different values of $\mu M$. This time we see cases ($\mu M=0.7$ and $\mu M=0.3$) for which the small $\mu$ approximation of equation~\eqref{testomega} fails. For the smaller values of $\mu M$ ($0.03$ and $0.003$), we were not able to extend the curves further into the test field region, so it is hard to tell whether equation~\eqref{testomega} will give a good approximation in those cases.
Figure~\ref{f:l2sigma} shows plots of $\sigma$ vs $A$ for different values of $\mu M$. Once again, we see errors, relatively larger this time, for small values of $\mu M$.
For results with $\ell=3$, which are qualitatively similar to those with $\ell=2$, we refer the reader to the appendix.

\FloatBarrier
\section{Conclusions} \label{s:conclusions}

We have presented new astrophysically relevant {approximate} solutions to the Einstein-Klein-Gordon equations, describing quasi-stationary systems that contain a black hole surrounded by a self-gravitating scalar field distribution. Although spherically symmetric, these objects are parametrized by an angular momentum number $\ell$, and therefore they have been named ``noble gravitational atoms''. 
Noble gravitational atoms can vary greatly in shape, size, mass, and lifetime, depending on the following parameters: the angular momentum number $\ell$, the scalar field mass $\mu$, the black hole mass $M$, and the scalar field amplitude $A$. 

One feature that differentiates $\ell>0$ from $\ell=0$ noble gravitational atoms is that the former admit solutions with a density maximum located far away from the black hole horizon. 
Additionally, they do not always exhibit a density spike at the horizon; sometimes, they show a small dip instead. Nonetheless, we have seen that in all cases, these spikes and dips have a negligible contribution to the mass distribution.
This behavior contrasts with the expectations for conventional cold dark matter and could, at least in principle, offer a way to discriminate between the two hypotheses.

We also find some similarities between the $\ell>0$ and the $\ell=0$ cases. 
Some solutions are extremely dilute, extremely large relative to the central black hole, and extremely long-lived, with lifetimes even larger than that of the Universe. 
Interestingly, these properties, together with the particular values of black hole and scalar field masses for which they exist, are compatible with ultralight dark matter in galactic cores.
Large, dilute, and long-lasting noble gravitational atoms with $\ell >0$ are also similar to those with $\ell=0$ in that they coincide almost perfectly with $\ell$-boson stars (of the same $\ell$), except only for an extremely small region close to the event horizon.

In general, noble gravitational atoms, as their $\ell$-boson star counterparts, 
can become larger as the angular momentum number $\ell$ increases. 
This characteristic may help extend the galactic halo core model mentioned earlier to cover a larger fraction of the galaxy, more so if one also considers solutions that combine multiple angular momentum numbers $\ell$ (and possibly even multiple mode numbers $n$ beyond the ground state), potentially allowing the model to reproduce the full galactic size.

{ In order to asses whether noble gravitational atoms can be regarded as viable astrophysical candidates, there is a series of open questions that remain to be addressed. A key aspect is their stability. $\ell$-boson stars are known to be stable against radial perturbations~\cite{Alcubierre:2019qnh, Alcubierre:2021mvs,Roque:2023sjl}, but they have been shown to be unstable against generic (non-radial) perturbations for $\ell>1$, at least in the nonrelativistic regime~\cite{Nambo:2023yut}. Similar behavior could be expected for noble gravitational atoms; however, a detailed stability analysis is required to establish this conclusively.}

{ Another important aspect of these new configurations concerns their gravitational wave emissions. Noble gravitational atoms are expected to exhibit gravitational spectra with distinctive features that differ from those of simple vacuum black hole solutions. The role of the angular momentum parameter $\ell$ has been explicitly studied in simulations of binary  $\ell$-boson star mergers~\cite{Jaramillo:2022zwg}, where it was shown to affect the emitted radiation. Such distinctive signatures may be observable with current and future gravitational wave detectors~\cite{Lira:2024cma}, potentially providing a mechanism to prove the existence of $\ell$ scalar field distributions. A detailed investigation of these signals is left for future work.}

{ Finally, we emphasize that the specific value $m\sim 10^{-22}\rm{eV}$ used for the results presented in our figures should be regarded purely as a reference choice. Current constraints on ultralight dark matter are in tension when different cosmological and astrophysical observations are combined (see, e.g., figure~18 in reference~\cite{Ferreira:2020fam}). On the one hand, cosmological probes such as the Lyman-alpha forest, the cosmic microwave background, and large-scale structure typically require masses of the order of $m\gtrsim 10^{-22} - 10^{-21}\rm{eV}$ if the ultralight field is to constitute the dominant component of dark matter, in order to avoid excessive suppression of small-scale structure~\cite{Lin:2023yso}. On the other hand, fits to the internal dynamics of dwarf spheroidal galaxies, which are dominated by dark matter, often favor lighter masses, $m\sim 10^{-22}\rm{eV}$, or even below, motivated by the ability of wave-like effects to alleviate the cusp profiles suggested by $N$-body simulations in the context of standard cold dark matter~\cite{Gonzalez-Morales:2016yaf}.}

{ As a consequence, there is currently no single mass range that simultaneously satisfies all observational constraints within the simplest realizations of ultralight dark matter, and the precise bounds remain highly model dependent. Moreover, many of these constraints rely on specific assumptions regarding the structure and equilibrium profiles of dark matter halos, with most galactic analyses focusing on the standard spherically symmetric $\ell=0$ ground-state configuration. Noble gravitational atoms, by contrast, correspond to distinct equilibrium states, and their density and kinematic profiles can differ substantially from those usually assumed. This opens the possibility that the corresponding phenomenological bounds on the boson mass may be modified, potentially reconciling cosmological and astrophysical constraints. Assessing this in detail would require dedicated fits to observational data using the profiles studied here, which lies beyond the scope of the present work.}

\ack{ This work was partially supported by SECIHTI-SNII,
by  PAPIIT-UNAM projects IN100523, IN110523 and IN107026, by CIC grant No.~18315 of the Universidad Michoacana de San Nicol\'as de Hidalgo,
by the programme HORIZON-MSCA2021-SE-01 Grant No.~NewFun-FiCO101086251.
A.D.T. acknowledges support from
DAIP project CIIC 2024 198/2024.  DN acknowledges the sabbatical support given by PAPIIT-UNAM
in the elaboration of the present work.
MM acknowledges financial support from CONICET (PIP 11220210100914CO).}
\newline

\appendix

\noindent {\bf Appendix}
\section{\texorpdfstring{$\ell=3$}{l=3} noble gravitational atoms}\label{app.ell=3}
\begin{figure}[tbh] \centering
\includegraphics[width=0.66\textwidth]{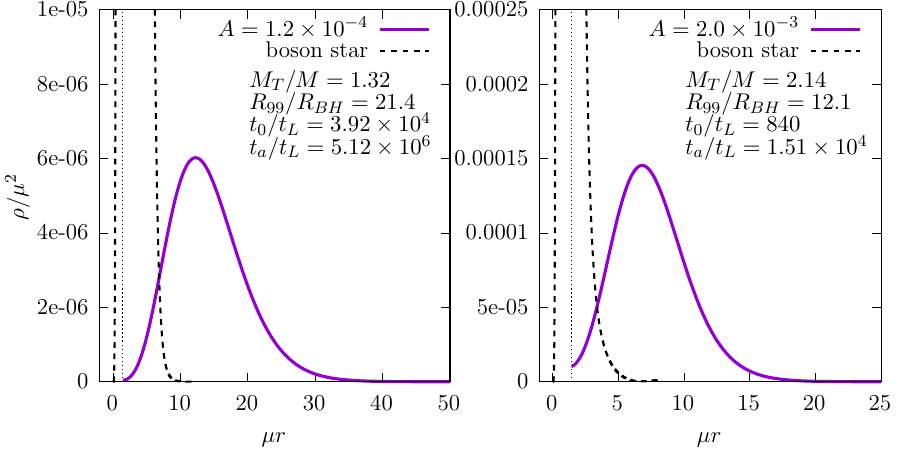}
    \caption{ \justifying
    Density $\rho$ of noble gravitational atoms with $\ell=3$, $\mu M=0.7$, and different amplitudes $A$.  \label{f:l3rho_comp}} 
\end{figure}

\begin{figure}[tbh] \centering
\includegraphics[width=0.45\textwidth]{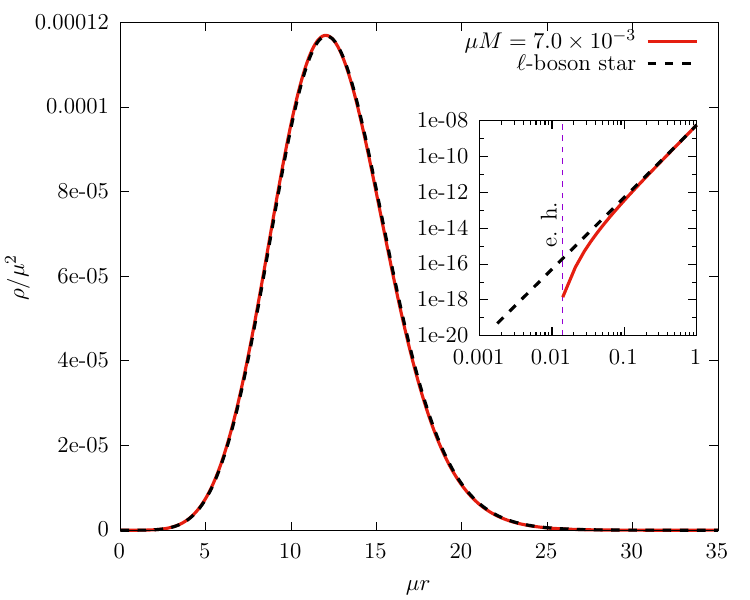}
\includegraphics[width=0.45\textwidth]{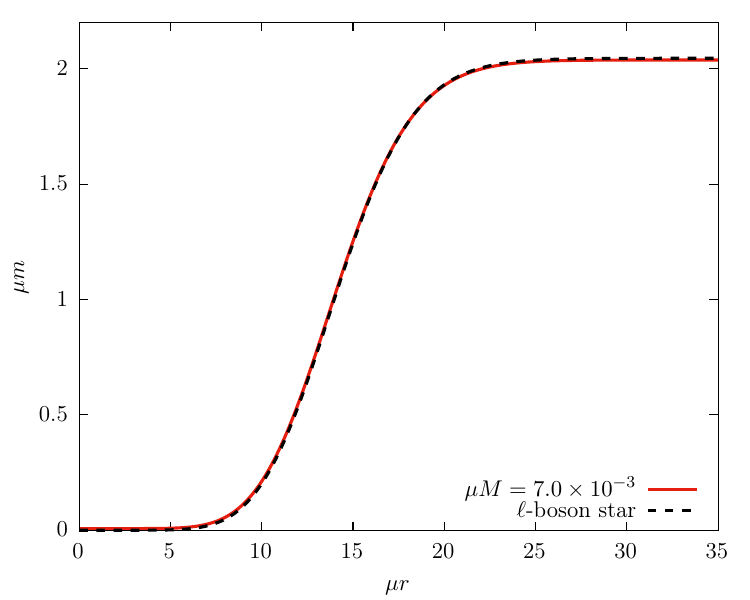}
    \caption{ \justifying
    Density $\rho$ and mass $m$ of noble gravitational atom and $\ell$-boson star with $\ell=3$ and $\mu M=7\times 10^{-3}$. The amplitudes are $\bar{u}_0=2.99\times 10^{-5}$ and $A$ such that the profiles coincide, according to equation~\eqref{u0tilde}. 
    \label{f:l3rho}} 
\end{figure}

\begin{figure}[tbh] \centering
\includegraphics[width=0.5\textwidth]{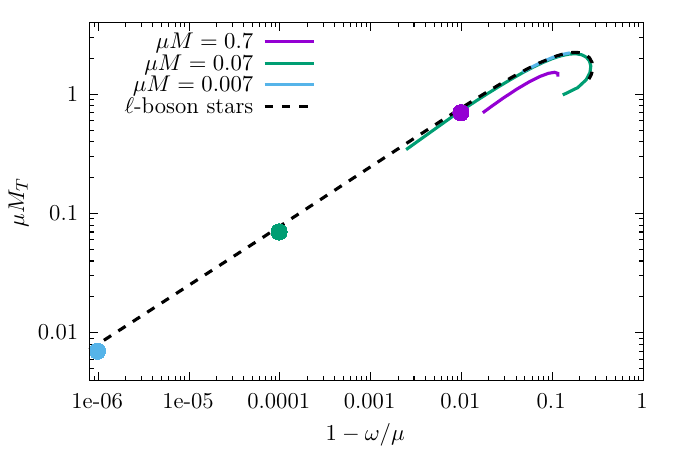}
    \caption{ \justifying
    $M_T$ vs $\omega$ of noble gravitational atoms and $\ell$-boson stars with $\ell=3$. 
    \label{f:l3Mw}} 
\end{figure}

\begin{figure}[tbh] \centering
\includegraphics[width=0.99\textwidth]{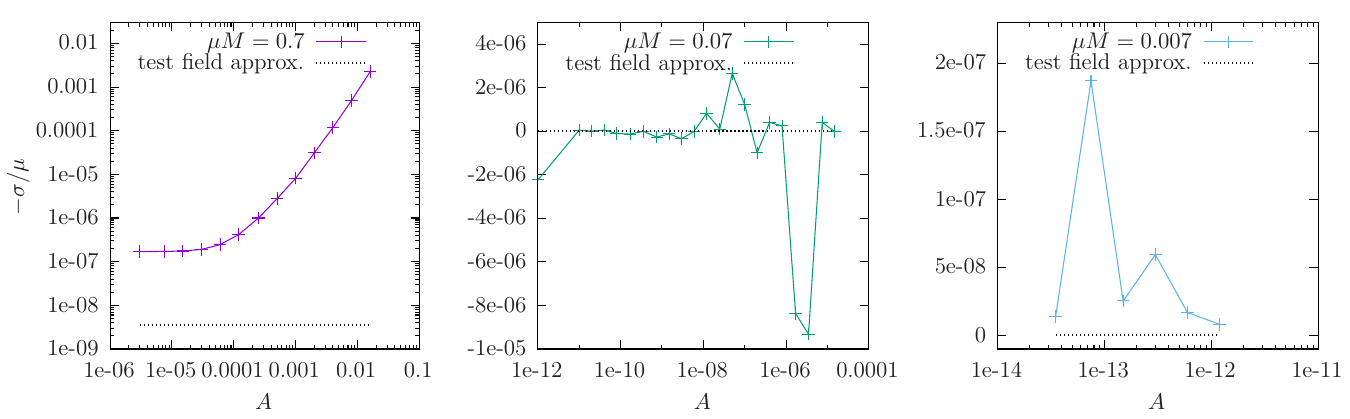}
    \caption{ \justifying
    $\sigma$ vs $A$ of noble gravitational atoms with $\ell=3$. 
    \label{f:l3sigma}} 
\end{figure}

In this appendix, we include results for $\ell=3$, which were not reported in the main text.
In figures~\ref{f:l3rho_comp} and~\ref{f:l3rho}, we show some density profiles.
Figure~\ref{f:l3Mw} shows the total mass $M_T$ vs real frequency $\omega$.
Figure~\ref{f:l3sigma} shows the imaginary frequency $\sigma$ vs amplitude $A$.
We find the equivalence
$A= 0.00366\, (2M)^3 u_0$,
so that $c_3=0.00366$. Note that the qualitative behavior of the $\ell=3$ case is similar to that of $\ell=2$, and a similar trend is also expected for all cases with $\ell\ge 2$ as is the case with $\ell$-boson stars. 

\FloatBarrier
\section{Collisionless cold dark matter halo model}\label{app.CDM}

In this appendix, we provide the details needed to reproduce the green curve of figure~\ref{f:rhospike}. In the collisionless cold dark matter model, the combined black hole dark matter halo system is commonly described using the following piecewise density profile~\cite{Bertone:2004pz,Ullio:2001fb,Lacroix:2018zmg}
\begin{align}
\rho_{\rm CDM}(r) =
\begin{cases}
0 & r < r_0,  \\
\rho_{\mathrm{NFW}}(R_{\mathrm{sp}}) \left( \dfrac{R_{\mathrm{sp}}}{r} \right)^{\gamma} & r_0 \leqslant r < R_{\mathrm{sp}}  \\
\rho_{\mathrm{NFW}}(r) & r \geqslant R_{\mathrm{sp}},
\end{cases}  ,
\label{rhocdm}
\end{align}
where each radial interval is determined by the astrophysical processes that dominate at that particular scale. 

In the region $r\ge R_{\textrm{sp}}$, where the influence of the black hole is negligible, we assume the Navarro-Frenk-White profile, 
\begin{equation}
\rho_{\mathrm{NFW}}(r) = \frac{\rho_{s}}{\left( \dfrac{r}{r_{s}} \right) \left( 1 + \dfrac{r}{r_{s}} \right)^{2}},
\label{rhodm}
\end{equation}
as motivated by cosmological $N$-body simulations~\cite{Navarro:1995iw,Navarro:1996gj}. In this expression, $\rho_s$ and $r_s$ represent a characteristic density and a scale radius, whose values depend on the specific galaxy under consideration. For the Milky Way~\cite{McMillan:2016jtx,deSalas:2020hbh}, we can take $r_{s}=18.6~\rm{Kpc}$ and $\rho_s=\left(\frac{R_0}{r_s}\right)\left(1+\frac{R_0}{r_s}\right)^2 \rho_\odot$, where $\rho_\odot=0.4~\rm{GeV~cm}^{-3}$ is the local dark matter density in the Solar System, located at a galactocentric radius of $R_0=8.2~\rm{Kpc}$. 

As the radius decreases and one approaches the central black hole, additional dynamical processes can modify the Navarro-Frenk-White profile in the region  $r_0\le r<R_{\rm{sp}}$, forming a black hole spike. This spike is commonly described by the phenomenological profile $\rho_{\textrm{CDM}}\propto(R_{\textrm{sp}}/r)^\gamma$, where $\gamma$ denotes the slope of the spike and $R_{\textrm{sp}}$ represents its characteristic radius. Under the assumption that the black hole grows adiabatically at the center of an initially cuspy halo, Gondolo and Silk estimated a spike slope of $\gamma=7/3$~\cite{Gondolo:1999ef}.
The spatial extension of the spike is much more uncertain, as it depends on the detailed evolutionary history of the black hole and its host. For practical purposes, we will use the conservative estimate $R_{\textrm{sp}}=10~\rm{pc}$, consistent with the choice made in~\cite{Ullio:2001fb}.
Finally, inside the event horizon, $r<r_0$, we have set the dark matter density to zero.

\FloatBarrier


\end{document}